\documentclass[preprint,12pt,authoryear]{elsarticle}

\usepackage{amssymb}
\usepackage{amsmath}
\usepackage{mathabx}
\usepackage{booktabs}
\usepackage{multirow}
\usepackage{graphicx}
\usepackage{hyperref}
\usepackage{xcolor}
\usepackage{caption}
\usepackage{lineno}

\newcommand{\appropto}{\mathrel{\mkern-20mu\vcenter{
  \offinterlineskip\halign{\hfil$##$\cr
    \propto\cr\noalign{\kern2pt}\sim\cr\noalign{\kern-2pt}}}\mkern-20mu}}

\newcommand{\showrevisions}{0}
\newif\ifshowrevisions
\showrevisionstrue   

\newcommand{\revised}[1]{%
    \if1\showrevisions%
        \textcolor{blue}{#1}%
    \else%
        #1%
    \fi%
}

\begin{document}
\begin{frontmatter}



\title{Learned Hemodynamic Coupling Inference in Resting-State Functional MRI} 

\author[label1]{William Consagra}
\author[label2]{Eardi Lila}
\affiliation[label1]{organization={Dept. of Statistics, University of South Carolina},
             addressline={1523 Greene St},
             city={Columbia},
             postcode={29225},
             state={SC},
             country={USA}}
             
\affiliation[label2]{organization={Dept. of Biostatistics, University of Washington},
             addressline={4333 Brooklyn Avenue NE},
             city={Seattle},
             postcode={98195},
             state={WA},
             country={USA}}
             
\begin{abstract}
Functional magnetic resonance imaging (fMRI) provides an indirect measurement of neuronal activity via hemodynamic responses that vary across brain regions and individuals. Ignoring this hemodynamic variability can bias downstream connectivity estimates. Furthermore, the hemodynamic parameters themselves may serve as important imaging biomarkers. Estimating spatially varying hemodynamics from resting-state fMRI (rsfMRI) is therefore an important but challenging blind inverse problem, since both the latent neural activity and the hemodynamic coupling are unknown. In this work, we propose a methodology for inferring hemodynamic coupling on the cortical surface from rsfMRI. Our approach avoids the highly unstable joint recovery of neural activity and hemodynamics by marginalizing out the latent neural signal and basing inference on the resulting marginal likelihood. To enable scalable, high-resolution estimation, we employ a deep neural network combined with conditional normalizing flows to accurately approximate this intractable marginal likelihood, while enforcing spatial coherence through priors defined on the cortical surface that admit sparse representations. Uncertainty in the hemodynamic estimates is quantified via a double-bootstrap procedure. The proposed approach is extensively validated using synthetic data and real fMRI datasets, demonstrating clear improvements over current methods for hemodynamic estimation and downstream connectivity \revised{reproducibility}. 
\end{abstract}




\end{frontmatter}

\section{Introduction}
\label{sec:introduction}
\subsection{Motivation}
Functional magnetic resonance imaging (fMRI) is a non-invasive imaging technique widely used to study neural activity. fMRI collects time-dependent blood oxygenation level–dependent (BOLD) signals, which reflect hemodynamic changes associated with neural activity \citep{Ogawa1992}. Accurate characterization of the hemodynamic coupling, i.e., the forward model that maps neural activity to BOLD signals, enables inference of underlying neural activity by solving the \textit{hemodynamic inverse problem} \citep{Buckner2003}. However, this problem is complicated by the fact that the hemodynamic coupling is partially unknown and varies both across brain regions and between individuals \citep{loh2008residual,Bailes2023}. Failure to account for the spatially varying hemodynamics can bias downstream estimates of functional connectivity  \citep{lindquist2009modeling,Rangaprakash2018,rangaprakash2023confound}. Additionally, the hemodynamic properties are themselves of independent interest as potential imaging biomarkers.
Indeed, systematic hemodynamic alterations have been identified in a variety of neuropsychiatric conditions, including autism \citep{YAN2018320}, schizophrenia and bipolar disorder \citep{Wenjing2021}, and obsessive compulsive disorder \citep{Rangaprakash2021}.
\par 
There exists a substantial body of work on estimating hemodynamic coupling in task/event-based fMRI designs \citep{Lange2002,wang2013multiscale,ZHANG20121754,ZHANG2013136}. In this setting, the neural signals are assumed to be known from the experimental design, making inversion considerably easier than in resting-state fMRI (rsfMRI), where both neural signals and hemodynamic response are unknown. The resting-state setting therefore leads to a challenging \textit{blind inverse problem}, which is the focus of this work.

\subsection{Related Work}
The hemodynamic coupling is often modeled as a linear time-invariant system (LTI), in which the BOLD signal is expressed as a convolution of the underlying neural signals and the \textit{hemodynamic response function} (HRF). In rsfMRI, many approaches avoid explicitly addressing the resulting blind inverse problem by assuming a known, spatially fixed HRF and then recover neural activity by solving the resulting (misspecified) linear inverse problem under a variety of spatio-temporal smoothness/sparsity priors \citep{KARAHANOGLU2013121,CABALLEROGAUDES2019116081,uruñuela2024}. In contrast, methods accommodating unknown hemodynamic couplings are less common: \cite{WU2013365} use a two-stage procedure that first thresholds to detect activation and then estimates the HRF; \cite{singh2020scalable} use a reparameterization approach based on the Wiener deconvolution and estimate the HRF using nonlinear least squares. Both approaches operate essentially \textit{pointwise}, that is, they estimate local hemodynamic coupling using only information from each spatial location (mesh vertex or atlas parcel) independently, and therefore do not exploit spatial dependence across the cortex. This is a notable limitation in rsfMRI, where signal-to-noise ratios are extremely low and borrowing spatial information can substantially improve estimation. An alternative approach from \cite{Cherkaoui2021} uses a template-shifted HRF and combines atlas-based low-rank and total-variation penalties for estimation. While this approach partially incorporates spatial dependence, it requires course atlas-based discretizations of the cortical surface, which severely limits the spatial resolution of the recovered hemodynamic properties.
\par 
A separate line of work uses nonlinear state-space models for coupling latent neural dynamics with more general, biophysically interpretable hemodynamic forward models \citep{buxton1998dynamics}, with inference procedures adapted to resting-state designs \citep{havlicek2011dynamic}. However, due to heavy computational costs, such approaches are generally limited to modeling a small number of brain regions \citep{cao2019functional}, effectively imposing a strong limitation on spatial resolution. Moreover, for many practical acquisition protocols, several of the hemodynamic parameters in such models have questionable statistical identifiability \citep{zayane2015sensitivity}.
\par 
There is an extensive body of work addressing the closely related problem of blind image deblurring, where the goal is to recover both an unknown sharp image and the corresponding blur kernel from a noisy, blurred observation. Most approaches estimate image and kernel jointly, via alternating optimization \citep{chan1998total} or deep learning based approaches \citep{Gossard2024}. In contrast, \cite{wipf2014revisiting} outlines that, while the joint reconstruction problem is often severely ill-posed, the task can be decomposed into two more manageable subproblems: first marginalizing the image out of the data likelihood and estimating the blur by maximizing the resulting marginal likelihood, and then performing non-blind deblurring given the estimated kernel. While related, this class of blind image deblurring problems is structurally different from the blind hemodynamic inverse problem, as the former typically involves only a single unknown image-kernel pair, whereas the latter involves many coupled blind problems with complex underlying spatial dependencies in both signals and hemodynamics. 

\subsection{Our Contribution}
In this work, we propose a novel methodology for estimating hemodynamic coupling in rsfMRI. In the spirit of \cite{wipf2014revisiting}, a key aspect of our approach is to avoid the highly unstable joint recovery of hemodynamics and neural activity by integrating out the latent neural signals and performing inference via a learned approximation to the marginal likelihood. This decouples estimation of the hemodynamic forward model from the ill-posed neural signal reconstruction. While this marginal likelihood is analytically intractable, we approximate it efficiently using a conditional neural spline normalizing flow trained on synthetic samples from a biologically motivated neural–hemodynamic simulator model. This simulator is calibrated to real fMRI signals using a spectral moment–matching approach that guarantees the generative process produces physiologically plausible signals. Another key aspect of our approach is the incorporation of spatial regularization based on a latent Gaussian process prior to promote locally similar hemodynamic couplings across the cortical surface. We propose an estimation algorithm that leverages automatic differentiation of the learned likelihood and the sparsity structure of the Gaussian process prior to enable computational scalability to ultra high-resolution surface meshes. To the best of our knowledge, this is the first approach to perform spatially regularized, global hemodynamic coupling inference directly on the cortical surface in rsfMRI. Uncertainty in the hemodynamic estimates is quantified via a double-bootstrap procedure, and automatic procedures for the selection of all model hyperparameters are proposed. Estimating the hemodynamic coupling enables the use of off-the-shelf methods to recover neural signals from BOLD data. Through extensive experiments using both synthetic and real data, we demonstrate substantial improvements over competing approaches in both hemodynamic estimation and downstream functional connectivity \revised{reproducibility}. Code for our method and scripts to reproduce the experimental results are publicly available: \url{github.com/Will-Consagra/surface-hemodynamic-inference.}

\section{Models and Background}\label{sec:models}

\begin{figure}
    \centering
    \includegraphics[width=\textwidth]{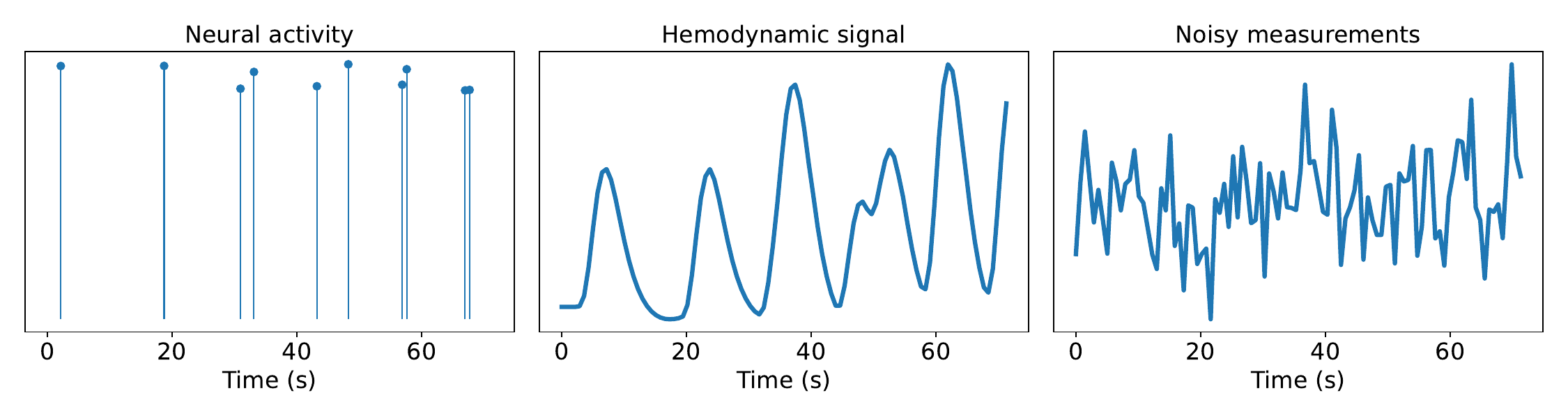}
    \caption{Simulation from the data generative model. (Left) Neural activity sampled from \eqref{eqn:signal_generative_model}. (Middle) Hemodynamic response obtained by convolving the neural activity with the HRF \eqref{eqn:double_gamma_hrf} under forward model \eqref{eqn:LTI_HR_model}. (Right) Simulated BOLD measurements generated according to \eqref{eqn:bold_forward_model} ($t_r=0.72\,\mathrm{s}$).}
    \label{fig:data_generative_model}
\end{figure}

This section outlines the models used for simulating neural activity and observed hemodynamic signals. Specifically, Section~\ref{ssec:model_neural_signals} introduces the neural activity model, Section~\ref{ssec:model_hemodynamics} describes the signal-induced hemodynamic response model, and Section~\ref{ssec:statistical_model} proposes a statistical model for the observed rsfMRI data. Figure~\ref{fig:data_generative_model} illustrates the generative process. 
\revised{Importantly, the inversion method in Section~\ref{sec:method} is simulation-based, so alternative neural signal, hemodynamic, or noise models can be accommodated by modifying the simulator and retraining the inversion model, while leaving the remainder of the inference procedure largely unchanged.} Throughout, we will assume the fMRI time series are sampled at a constant repetition time $t_{r}>0$, with accompanying acquisition times $t_m=(m-1)t_{r}$, for $m=1,...,M$, spanning the interval $[0,(M-1)t_{r}]$.

\subsection{Modeling Neural Population Dynamics}\label{ssec:model_neural_signals}

Denote the brain cortical surface by $\Omega\subset\mathbb{R}^{3}$. At location $x\in\Omega$, we model the time-dependent neural activity, denoted by $s(x,t)$, as a sum of scaled and shifted Dirac delta functions of the form
\begin{equation}\label{eqn:shiftedSpike}
    s(x,t) = \sum_{i=1}^{n_{x}} a_{x,i}\delta(t-t_{x,i}). 
\end{equation}
We assume that the number of spikes $n_{x}$ and their occurrence times $t_{x,1}, ...,t_{x,n_{x}}$ are generated by a homogeneous Poisson process on $[0,(M-1)t_{r}]$ with rate $\lambda_{x}$. The spike amplitude parameters $\{a_{x,i}\}_{i=1}^{n_{x}}$ are assumed to be independent and identically distributed uniform random variables on the interval $(a_{\min},a_{\max})$. Given the local heterogeneity in grey-matter tissue composition, along with the complex network of non-local white-matter structural connections, the rate $\lambda_{x}$ is expected to vary across $\Omega$. To capture the marginal effect of this variability, we assume $\lambda_{x}$ is a random uniform variable on the interval $(\lambda_{\min},\lambda_{\max})$. Putting this all together, we have the following hierarchical model for neural activity at $x\in\Omega$:    
\begin{equation}\label{eqn:signal_generative_model}
    \begin{aligned}
        &\lambda_{x} \sim \text{Unif}([\lambda_{min}, \lambda_{max}]) \\
        &(t_{x,1},t_{x,2}, ..., t_{x,n_{x}}) \sim \text{HPP}(\lambda_{x} t_r(M-1) ) \\
        &a_{x,i} \sim\text{Unif}([a_{min},a_{max}]), \quad i=1,...,n_{x}\\
        &s(x,t) = \sum_{i=1}^{n_{x}} a_{x,i}\delta(t-t_{x,i})
    \end{aligned}
\end{equation}
    

\subsection{Modeling the Hemodynamic Response}\label{ssec:model_hemodynamics}

Changes in local neural activity induce changes in cerebral blood flow. We model this \textit{hemodynamic coupling} using an operator $\mathcal{H}_{\theta}:L^2(\mathbb{R})\mapsto\mathcal{C}(\mathbb{R})$, parameterized by $\theta\in\Theta\subset\mathbb{R}^{J}$. Due to tissue heterogeneity, the hemodynamic coupling can vary across $\Omega$. We accommodate this variability by modeling the hemodynamic parameters $\theta$ as a $J$-dimensional field on the cortical surface, $\theta:\Omega\mapsto\Theta$. Therefore, at $x\in\Omega$, the hemodynamic response $\mathcal{H}_{\theta(x)}$ is fully characterized by the J-dimensional parameter $\theta(x)$.
\par 
In this work, we will consider LTI forward models of the form
\begin{equation}\label{eqn:LTI_HR_model}
 \mathcal{H}_{\theta(x)}[s(x,\cdot)](t) = \int_{0}^th_{\theta(x)}(t-u)s(x,u)du,
\end{equation}
where $h_{\theta(x)}$ is a parametric kernel. The hemodynamic parameters are often constrained componentwise to ensure biophysically plausible behavior, so we assume that the parameter space is 
$\Theta = \{\theta\in\mathbb{R}^{J}: \theta_{j,\min}\le \theta_j\le \theta_{j,\max}\}$. While specifying a reasonable prior model for the dependence structure of $\theta$ over $\Omega$ is challenging, we can nevertheless simulate marginal hemodynamics at $x\in\Omega$ by adopting a non-informative box-uniform prior over the parameter space, $\theta(x) \sim \text{Unif}(\Theta)$. 
\par 
For estimation, it will be useful to map the constraints on $\Theta$ to an unconstrained space. To do so, we define the modified probit link function $g(\theta)=(g_1(\theta_1),...,g_J(\theta_J))$ by
$$
\tilde{\theta}_j := g_j(\theta_j) = \Phi^{-1}\left(\frac{\theta_j-\theta_{j,\min}}{\theta_{j,\max} - \theta_{j,\min}}\right), \quad j=1,...,J
$$ 
where $\Phi^{-1}$ denotes the inverse cdf of the standard normal distribution. We denote the transformed parameters 
$\tilde{\theta}=g(\theta)$ and transformed parameter space by $\tilde{\Theta} = \{\tilde{\theta}=g(\theta):\theta\in\Theta\}$. 

\subsection{Statistical Model for Observed Data}\label{ssec:statistical_model}

For a set of cortical locations $\mathcal{X}:=\{x_1,...,x_{V}\}\subset\Omega$, we gather noisy observations of the induced hemodynamic response at each of the acquisition times $t_1,...,t_{M}$. The noise process contaminating the signals includes physiological and scanner contributions that may induce temporal autocorrelation \citep{lindquist2008statistical}. For sufficiently small $t_r$, this autocorrelation can be modeled and removed during preprocessing, yielding an approximately white error process, which we assume here. The resulting Gaussian measurement noise model for the observed fMRI signals is then given by
\begin{equation}\label{eqn:bold_forward_model}
    y_{v,m} = \mathcal{H}_{\theta(x_{v})}[s(x_v,\cdot)](t_m) + \epsilon(x_v,t_m),
\end{equation}
for $v=1,...,V;\, m=1,...,M$, with $\epsilon$ a stationary zero mean white noise process with $\text{Cov}(\epsilon(x,t), \epsilon(x^{\prime},t^{\prime})) = \sigma^2\mathbb{I}\{x=x^{\prime},t=t^{\prime}\}$.

\section{Methodology}\label{sec:method}
\begin{figure}[!t]
    \centering
    \includegraphics[width=\textwidth]{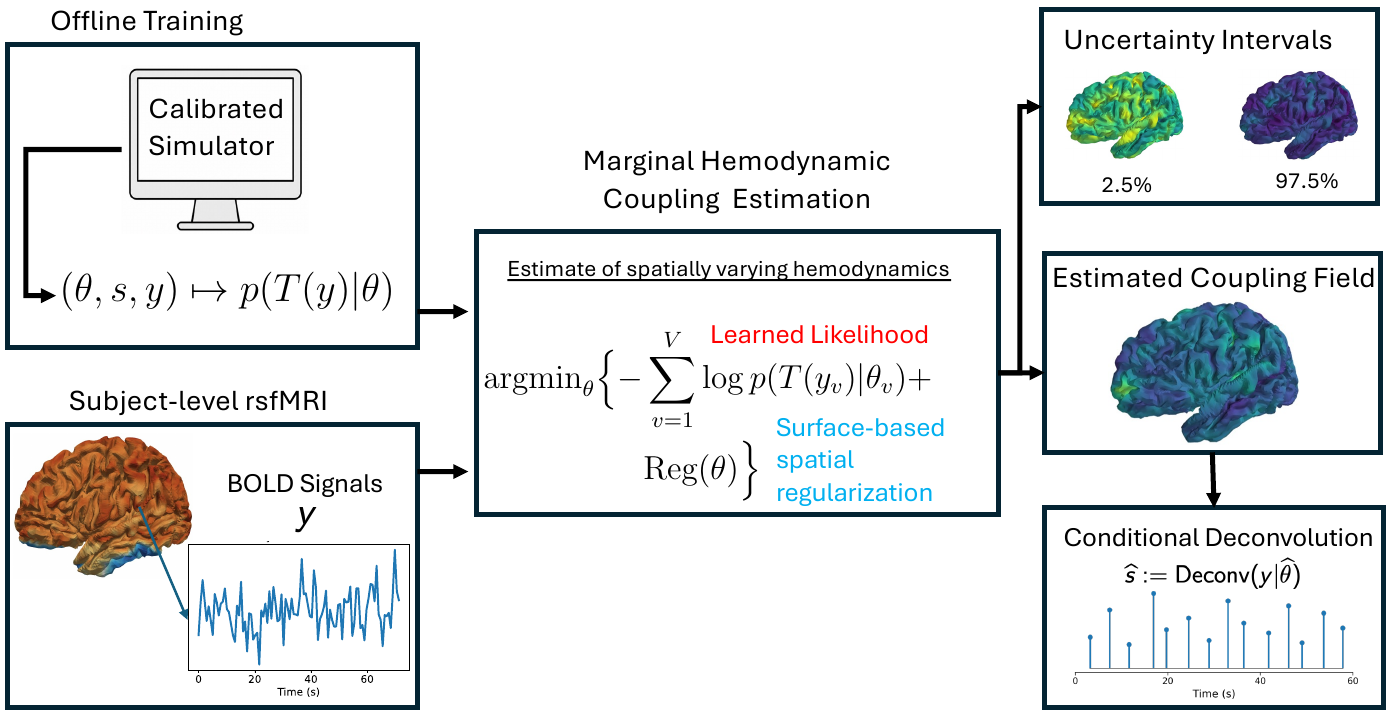}
    \caption{\revised{Methodology overview. 
    A deep generative model is trained on forward simulations from a simulator, calibrated to match the spectral properties of real rsfMRI signals, to approximate the target intractable marginal likelihood for hemodynamic coupling. This learned likelihood is combined with a Gaussian process prior on the cortical surface to estimate the spatially varying hemodynamic coupling field. Estimation uncertainty is quantified using a resampling procedure with a calibration step to accommodate potential bias induced by the prior regularizer. Optional second-stage signal deconvolution is performed conditionally on the estimated hemodynamic coupling field.}}
    \label{fig:pipeline}
\end{figure}
As outlined in Section~\ref{sec:introduction}, direct joint estimation of the neural signals and hemodynamic parameters is often severely ill-posed and can lead to unstable inference.  
\revised{We instead marginalize over the neural-signal generative process and base hemodynamic coupling estimation on the resulting marginal likelihood. The overall inference pipeline is summarized in Figure~\ref{fig:pipeline}. Since this marginal likelihood is analytically intractable, we use a simulation-based inference approach \citep{cranmer2020frontier} and approximate it using a deep generative model trained on synthetic fMRI data (Section~\ref{ssec:learnedLik}). To share information across space and regularize the inverse problem, we impose a Gaussian process prior on the hemodynamic parameter field (Section~\ref{ssec:spatial_prior}). We then derive a scalable estimation algorithm for marginal hemodynamic coupling estimation that leverages automatic differentiation through the learned likelihood together with the sparsity of the surface-based spatial regularizer to enable inference on high-resolution cortical surface meshes (Section~\ref{ssec:point_estimation}). Data-driven procedures for automatic selection of all model hyperparameters are provided in Section~\ref{ssec:hyperparam_select}. To quantify uncertainty in the estimated hemodynamic coupling, we propose a resampling procedure that preserves temporal dependence in the observed fMRI signals (Section~\ref{ssec:uq}). Finally, conditional on the estimated hemodynamic coupling, neural signal recovery reduces to a non-blind inverse problem. In Section~\ref{ssec:signal_deconv}, we outline a default signal deconvolution method aligned with the neural-signal prior introduced in Section~\ref{ssec:model_neural_signals}.} 
\par 
We denote the data matrix $Y\in\mathbb{R}^{V\times M}$, where the $v$-th row is the vector $y_v=(y_{v,1},...,y_{v,M})$, with $y_{v,m}$ defined by Equation~\eqref{eqn:bold_forward_model}. With slight abuse notation, define $\boldsymbol{\theta},\tilde{\boldsymbol{\theta}}\in\mathbb{R}^{V\times J}$ to be the hemodynamic parameter field, and its transformation under $g$, discretized over $\mathcal{X}$; i.e., $\boldsymbol{\theta}_{v,j}=\theta_j(x_v)$ and $\tilde{\boldsymbol{\theta}}_{v,j} = \left( g(\boldsymbol{\theta}) \right)_{v,j}$, respectively. Finally, define $\text{vec}:\mathbb{R}^{n\times m}\mapsto\mathbb{R}^{nm}$ as the vectorization operator which stacks columnwise. 

\subsection{Learned Marginal Likelihood}\label{ssec:learnedLik}
We define the marginal likelihood at location $x_{v}\in\mathcal{X}$, in terms of the unconstrained parameters $\tilde{\theta}=g(\theta)$, as
\begin{equation}\label{eqn:marginal_likelihood}
    p(y_v|\tilde{\theta}(x_{v})) = \int_{\mathcal{S}}p(y_{v}|s,\tilde{\theta},x_{v})p(s|x_{v})ds,
\end{equation}
where the likelihood $p(y_{v}|s,\tilde{\theta},x_{v})$ is defined by \eqref{eqn:bold_forward_model}, the point-wise signal prior $p(s|x_v)$ is defined by \eqref{eqn:signal_generative_model}, and $\mathcal{S}$ denotes the space of signals spanned by the generative model. For tractability, we ignore the complex spatial dependence among neural signals and instead we approximate the joint prior over $\mathcal{X}$ by a product of the marginal priors outlined in Section~\ref{ssec:model_neural_signals}. \revised{We emphasize that this assumption is introduced solely as a computational approximation to facilitate tractable inference and is likely to be violated in practice. We directly assess robustness to this misspecification in the synthetic-data experiments of Section~\ref{ssec:synth_results}.
Under this approximation,} it follows that  $p(Y|\theta, \mathcal{X}) \approx \prod_{v=1}^Vp(y_v|\tilde{\theta}(x_{v}))$, hence it suffices to focus on the location-specific model in \eqref{eqn:marginal_likelihood}. Two major challenges arise. First, the marginalization integral in  \eqref{eqn:marginal_likelihood} does not admit a closed-form solution. Although a Monte Carlo approximation can be obtained by repeatedly simulating from \eqref{eqn:shiftedSpike} and \eqref{eqn:bold_forward_model}, this is computationally prohibitive for large $V$, so a more scalable approximation is required. Second, $p(y_v|\tilde{\theta}(x_v))$ is typically high-dimensional, making it challenging to estimate. The remainder of this section outlines our approach to addressing these issues. For notational simplicity, we henceforth suppress the explicit dependence of \eqref{eqn:marginal_likelihood} on $x$. 
\par 
To handle the high-dimensionality of the observed signals $y$, we propose performing dimension reduction by learning a summary statistic $T:\mathbb{R}^{M}\mapsto\mathbb{R}^{D}$, where $D\ll M$, and basing inference off $ p(T(y)|\tilde{\theta})$  \revised{rather than the raw high-dimensional marginal likelihood $p(y|\tilde{\theta})$. If $T$ is a sufficient statistic, the summary marginal likelihood contains all information in $y$ relevant for inference on $\tilde{\theta}$, otherwise we can expect some information loss. In practice, outside of special cases such as exponential families, identifying a sufficient statistic is challenging, and there is no guarantee that $D\ll M$. Instead, we look for a $T$ that balances retaining enough information for inferring $\tilde{\theta}$ while providing sufficient dimension reduction, i.e., providing $D$ small enough for tractable conditional density approximation.} While a variety of candidates for such a $T$ exist, in this work we adopt the simple approach of \cite{jiang2017learning}, which targets the posterior mean
\begin{equation}\label{eqn:posterior_mean_summary}
    T(y) = \mathbb{E}_{p(\tilde{\theta}|y)}\left[\tilde{\theta}\right].
\end{equation}
The posterior mean has several attractive properties for serving as a summary statistic in our setting: i) it imposes that $D=J$, where $J$ is the dimension of $\tilde{\theta}$, ensuring significant dimension reduction, and ii) it can be estimated using standard $l_2$ risk minimization. 
\par 
Let $T_{\psi}:\mathbb{R}^{M}\mapsto\mathbb{R}^{J}$ be a flexible model depending on parameters $\psi$, and let  $p_{\gamma}(\cdot|\cdot):\mathbb{R}^{J}\mapsto\mathbb{R}^{J}$ be a flexible conditional density estimator depending on parameters $\gamma$. We approximate the target density $p(T(y)|\tilde{\theta})$ in two stages. We first learn the summary statistics by minimizing the empirical expectation 
\begin{equation}\label{eqn:posteriorSummaryNetLearn}
    \widehat{\psi}_{N} = \min_{\psi}\frac{1}{N}\sum_{i=1}^N \left\|T_{\psi}(y_i) - \tilde{\theta}_i \right\|_{2}^2, \quad (y_i,\tilde{\theta}_i)\sim p(\tilde{\theta}, y),
\end{equation}
\revised{where $N$ is the number of training data drawn from the simulation model.} Next, fixing $T_{\widehat{\psi}_{N}}$, we learn an approximation to the target marginal likelihood by maximizing the log-density 
\begin{equation}\label{eqn:likelihoodEmulator}
    \widehat{\gamma}_{N} = \max_{\gamma}\frac{1}{N}\sum_{i=1}^N\log p_{\gamma}(T_{\widehat{\psi}_{N}}(y_i)|\tilde{\theta}_i) \quad (y_i,\tilde{\theta}_i)\sim p(\tilde{\theta}, y).
\end{equation}
The training pairs $(y_i,\tilde \theta_i)$ from the marginal joint density $p(y,\tilde{\theta})=p(y|\tilde{\theta})p(\tilde{\theta})$ needed to form \eqref{eqn:posteriorSummaryNetLearn} and \eqref{eqn:likelihoodEmulator} can be obtained from the simulation model outlined in Section~\ref{sec:models}. 
\par 
In practice, we parameterize $T_{\psi}$ as a deep multilayer perceptron (MLP) with ReLU activations and $p_{\gamma}$ as a multilayer normalizing flow. More details on the network architecture and training can be found in Section~\ref{ssec:implementation_details}. 

\subsection{Prior Distribution}\label{ssec:spatial_prior}
Due to the spatial organization of cortical vasculature, hemodynamic parameters $\theta$, or equivalently $\tilde{\theta}$, are expected to vary smoothly across the cortex. Accordingly, we place a multivariate mean-zero Gaussian process prior on the transformed field, $\tilde{\theta}\sim GP(0,\mathcal{C}_{\tilde{\theta}})$, where the covariance operator is assumed to be diagonal $\mathcal{C}_{\tilde{\theta}} = \bigoplus_{j=1}^J \mathcal{C}_j$. That is, the $j$-th component of $\tilde{\theta}$ is modeled as an independent mean-zero GP with covariance operator $\mathcal{C}_j$. The remainder of this section details the model for the $j$-th component, thereby determining the full process.
\par 
On Euclidean domains, the Mat\'ern processes are a popular choice for modeling spatial fields due to their flexibility and the closed-form of their covariance kernel. Following \cite{lindgren2011explicit}, such processes can be generalized to spatial signals located on the highly non-linear cortical surface $\Omega$ as the weak solutions to the following stochastic partial differential equation 
\begin{equation}\label{eqn:MaternSPDE}
    \left(-\Delta_{\Omega} + \kappa_j^2 I\right)^{\beta_j/2}\tilde{\theta}_j(x) = \frac{1}{\tau_j} u(x), \qquad x \in \Omega
\end{equation}
with $\Delta_{\Omega}$ the Laplace-Beltrami operator on $\Omega$ and $u$ a white-noise process on $\Omega$, along with hyper-parameters $\beta_j>0$ dictating the smoothness, $\kappa_j>0$ the inverse bandwidth, and the marginal variance $\eta_j=\frac{\Gamma(\beta_j-1)}{\Gamma(\beta_j)4\pi\kappa_j^{2(\beta_j-1)}\tau_j^2}$ for $\tau_j >0$. Here, we fix $\beta_j=2$ for all $j$, as is commonly done.
\par
Assume the locations $x_1, \ldots, x_V$ coincide with the vertices of the triangulated cortical surface $\Omega$, as is common in surface-based representation of fMRI. Using the linear finite element basis $\{\phi_v\}_{v=1}^V$ induced by this triangulation \citep{lai2007spline}, define the inner product matrices with element-wise definitions 
\begin{equation}\label{eqn:mass_stiffness_matrices}
         C_{ij} = \int_{\Omega}\phi_i\phi_j\qquad G_{ij} = \int_{\Omega}[\nabla_{\Omega}\phi_i]^{\intercal}\nabla_{\Omega}\phi_j.
\end{equation}
Then it can be shown \revised{\citep{lindgren2011explicit}} that the weak solution to \eqref{eqn:MaternSPDE} under this basis implies that the discretized field $\tilde{\boldsymbol{\theta}_j}=\left(\tilde \theta_j(x_1),...,\tilde \theta_j(x_V)\right)^\intercal$ is normally distributed with precision matrix
$$
\begin{aligned}
    Q_j &= \tau_j^2\left(\kappa^2_j C + G\right)^{\intercal}C^{-1}\left(\kappa^2_j C + G\right).
\end{aligned}
$$
\revised{For completeness, Supplemental Section~\ref{supp:derivations} provides a derivation of this result.}
In practice, we use the so-called mass-lumping technique and approximate $C$ with a diagonal matrix. This diagonal approximation ensures that $Q_j$ is a sparse matrix. Together with the assumed independence across components, we have that $\text{vec}(\tilde{\boldsymbol{\theta}}^{\intercal})\sim \mathcal{N}(0, Q^{-1})$, where $Q=\text{BlockDiag}(Q_{1},...,Q_{J})$.

\subsection{Point Estimation}\label{ssec:point_estimation}

We estimate the hemodynamic parameters of the model outlined in Section~\ref{sec:models} by maximum a posteriori (MAP) estimation, combining the learned marginal likelihood from Section~\ref{ssec:learnedLik} with the spatial prior from Section~\ref{ssec:spatial_prior}. Specifically, we aim to maximize the approximate posterior density
\begin{equation}\label{eqn:posteriorlearnedSummary}
\begin{aligned}
    p(\tilde{\boldsymbol{\theta}}|Y)&\approx p(\tilde{\boldsymbol{\theta}}|T(y_1),...,T(y_V)) \appropto \left[\prod_{v=1}^Vp_{\widehat{\gamma}}(T_{\widehat{\psi}}(y_v)|\tilde{\theta}_v)\right]p(\tilde{\theta}_1, ..., \tilde{\theta}_V),
\end{aligned}
\end{equation} 
where $p(\tilde{\theta}_1, ..., \tilde{\theta}_V)$ denotes the spatial prior. We leverage automatic-differentiation of the learned marginal likelihood and the sparsity structure induced by the spatial prior to compute the MAP estimates efficiently.
\par 
Specifically, we compute an approximate MAP estimate using Newton’s method applied to the negative log-posterior
\begin{equation}\label{eqn:negLogPost}
\begin{aligned}
        -\log p(\tilde{\boldsymbol{\theta}}|Y) &\appropto - \sum_{v=1}^V\log p_{\widehat{\gamma}}(T_{\widehat{\psi}}(y_v)|\tilde{\theta}_v) + \frac{1}{2}\text{vec}(\tilde{\boldsymbol{\theta}})^{\intercal} Q\text{vec}(\tilde{\boldsymbol{\theta}}). 
\end{aligned}
\end{equation}
Newton's method forms a sequence of iterates $k=0,1,...$ defined by
\begin{equation}\label{eqn:NewtonIteration}
    \text{vec}(\tilde{\boldsymbol{\theta}}^{(k+1)}) = \text{vec}(\tilde{\boldsymbol{\theta}}^{(k)}) - \alpha_k\revised{w}^{(k)},
\end{equation}
where the update direction $\revised{w}^{(k)}$ solves the linear system
\begin{equation}\label{eqn:NewtonLinSystem}
    H(\text{vec}(\tilde{\boldsymbol{\theta}}^{(k)}))\revised{w}^{(k)} = G(\text{vec}(\tilde{\boldsymbol{\theta}}^{(k)})),
\end{equation}
with $G(\text{vec}(\tilde{\boldsymbol{\theta}}))\in\mathbb{R}^{VJ}$ and $H(\text{vec}(\tilde{\boldsymbol{\theta}}))\in\mathbb{R}^{VJ\times VJ}$ denoting the gradient and Hessian of \eqref{eqn:negLogPost}, respectively, and $\alpha_k$ the step-size. Although the dimensionality $VJ$ can be very large, the linear system in \eqref{eqn:NewtonLinSystem} is tractable due to its sparsity. Specifically, notice that the gradient has the form
\begin{equation}\label{eqn:logPostGrad}
\begin{aligned}
    G(\text{vec}(\tilde{\boldsymbol{\theta}})) &=  -\begin{pmatrix}
    \frac{\partial}{\partial\tilde{\theta}_{11}}\log p_{\widehat{\gamma}}(T_{\widehat{\psi}}(y_1)|\tilde{\theta}_{1}) \\
    \frac{\partial}{\partial\tilde{\theta}_{21}}\log p_{\widehat{\gamma}}(T_{\widehat{\psi}}(y_2)|\tilde{\theta}_{2}) \\
    \vdots\\
    \frac{\partial}{\partial\tilde{\theta}_{VJ}}\log p_{\widehat{\gamma}}(T_{\widehat{\psi}}(y_v)|\tilde{\theta}_{V}) 
    \end{pmatrix} + 
    Q\text{vec}(\tilde{\boldsymbol{\theta}}),
\end{aligned}
\end{equation}
and the Hessian decomposes as
\begin{equation}\label{eqn:logPostHess}
\begin{aligned}
    H(\text{vec}(\tilde{\boldsymbol{\theta}})) & = -\nabla_{\text{vec}(\tilde{\boldsymbol{\theta}})}^{2}\!
            \log \prod_{v=1}^{V} p_{\widehat{\gamma}}\!\bigl(T_{\widehat{\psi}}(y_{v}) \mid \tilde{\theta}_{v}\bigr) + Q,
\end{aligned}
\end{equation}
with 
\begin{equation}\label{eqn:obsInformation}
\begin{aligned}
&\Bigl[-\nabla_{\text{vec}(\tilde{\boldsymbol{\theta}})}^{2}\!
        \log \prod_{v=1}^{V} p_{\widehat{\gamma}}\!\bigl(T_{\widehat{\psi}}(y_{v}) \mid \tilde{\theta}_{v}\bigr)
 \Bigr]_{ii'}
 = \begin{cases}
   -\displaystyle\frac{\partial^{2}}
       {\partial\tilde{\theta}_{vj}\,\partial\tilde{\theta}_{vj'}}
       \log p_{\widehat{\gamma}}\!\bigl(T_{\widehat{\psi}}(y_{v}) \mid \tilde{\theta}_{v}\bigr),
       & i,i^{\prime} \in \mathcal{I}(v,j,j^{\prime})\\[10pt]
   0, & \text{otherwise}.
 \end{cases}
\end{aligned}
\end{equation}
Here, $\mathcal{I}$ is defined as
$$
\begin{aligned}
    \mathcal{I}(v,j,j^{\prime}) = &\{(i,i') \text{ such that } i=(j-1)V + v; \\ &i^{\prime}=(j^{\prime}-1)V + v;
    j,j^{\prime}\in \{1,...,J\};1\le v \le V\}.
\end{aligned}
$$
Notice that the element-wise definition \eqref{eqn:obsInformation} implies $\nabla_{\text{vec}(\tilde{\boldsymbol{\theta}})}^{2}\!\log \prod_{v=1}^{V} p_{\widehat{\gamma}}\!\bigl(T_{\widehat{\psi}}(y_{v}) \mid \tilde{\theta}_{v}\bigr)$ has only $J^2V$ non-zero terms. Given that the matrices $Q_{\kappa_{j},\tau_{J}}$ are sparse, it follows that the Hessian \eqref{eqn:logPostHess} is the sum of two sparse matrices and is therefore sparse. Moreover, both \eqref{eqn:logPostGrad} and \eqref{eqn:logPostHess} depend on partial derivatives of the learned marginal likelihood, which can be computed efficiently via back-propagation. Therefore, the linear system \eqref{eqn:NewtonLinSystem} can be solved efficiently using (preconditioned) conjugate gradient type algorithms. 
\par 
Given that the summary network $T_{\widehat{\psi}}(\cdot)$ targets the point-wise posterior mean \eqref{eqn:posterior_mean_summary}, we use the estimates $(T_{\widehat{\psi}}(y_v))_{v}$ to form the initialization $\text{vec}(\tilde{\boldsymbol{\theta}}^{(0)})$. The step-sizes $\alpha_k$ are updated at each iteration by performing an Armijo line search \citep{armijo1966minimization}.

\subsection{Hyperparameter Selection}\label{ssec:hyperparam_select}

\subsubsection{Calibration of Smoothness Prior}\label{sssec:hyperparam_smoothness}

The spatial prior precision matrices $(Q_j)_j$ depend on unknown hyperparameters $\kappa_j$ and $\tau_j$, controlling the bandwidth and marginal variance, respectively. To select these hyperparameters, we maximize a Laplace approximation to the log marginal likelihood $p(Y|\{\kappa_{j},\tau_{j}\}_j)$ around the MAP estimate, which has the following well known analytic form \revised{\citep{williams2006gaussian}}
\begin{equation}\label{eqn:margLik}
    \begin{aligned}
    \log p(Y|\{\kappa_{j},\tau_{j}\}_j) &\approx \sum_{v=1}^V\log p_{\widehat{\gamma}}(T_{\widehat{\psi}}(y_v)|\widehat{\tilde{\theta}}_v) - \frac{1}{2}\text{vec}(\widehat{\tilde{\boldsymbol{\theta}}})^{\intercal} Q\text{vec} (\widehat{\tilde{\boldsymbol{\theta}}}) \\
    & +\frac{1}{2}\log\det|Q| - \frac{1}{2}\log\det| H(\text{vec}(\widehat{\tilde{\boldsymbol{\theta}}}))|,
    \end{aligned}
\end{equation}
where $\text{vec}(\widehat{\tilde{\boldsymbol{\theta}}})$ is the MAP estimator formed in Section~\ref{ssec:point_estimation} and $H(\text{vec}(\widehat{\tilde{\boldsymbol{\theta}}}))$ is \eqref{eqn:logPostHess} evaluated at the MAP, which depends on the hyperparameters of interest  $\{\kappa_{j},\tau_{j}\}_j$ through $Q$. 
\par 
In the absence of additional prior information, we assume $\kappa_j=\kappa$ and $\tau_j=\tau$, for all $j$, to reduce the dimension of the hyperparameter search. Within this reduced space, we perform a grid search, selecting $\kappa,\tau$ that maximize \eqref{eqn:margLik} over a discrete set of locations.

\subsubsection{Calibration of Simulator}\label{sssec:simulator_calibration}

While $\theta_{j,\min}, \theta_{j,\max}$ are assumed known a-priori, the models in Section~\ref{sec:models} depend on several unknown simulator hyperparameters $\nu :=(\lambda_{\min},\lambda_{\max}, a_{\min}, a_{\max}, \sigma^2)$. To calibrate $\nu$ to our observed fMRI signals, we take a spectral moment-matching based approach. Specifically, let $p(\theta,s,\epsilon|\nu)$ denote the joint distribution induced by the simulator given $\nu$, and let $p^{*}(y)$ denote the true marginal distribution of the observed fMRI signals. We propose the following spectral moment-based loss for calibration 
\begin{equation}\label{eqn:moment_matching}
\begin{aligned}
        l_{s}(\nu) &= \sum_{\omega\in\mathcal{W}}\Big|\mathbb{E}_{p^{*}(y)}\Big[\mathcal{P}_{y}(\omega)\Big]- \mathbb{E}_{p(s,\theta,\epsilon|\nu)}\Big[\mathcal{P}_{\mathcal{H}_{\theta}[s](t) + \epsilon(t)}(\omega)\Big]\Big|,
\end{aligned}
\end{equation}
where $\mathcal{P}_{r}$ denotes an estimate of the power spectral density of the signal $r$ and
\begin{equation}\label{eqn:nyquist_frequencies}
    \revised{\mathcal{W}=\{\omega: \omega=\frac{k}{Mt_{r}};k\in\{-\frac{M}{2},...,\frac{M}{2}-1\}\}}
\end{equation} 
denotes the set of Nyquist frequencies. 
\par 
In practice, none of the quantities in \eqref{eqn:moment_matching} are available analytically. However, we can easily form an approximation via simulations, where $\mathbb{E}_{p^{*}(y)}$ is approximated using the observed BOLD signals and $\mathbb{E}_{p(s,\theta,\epsilon|\nu)}$ is approximated using forward model simulations. This results in the empirical loss function 
\begin{equation}\label{eqn:sim_based_moment_matching}
\begin{aligned}
        & l_{s}(\nu)  \approx \sum_{\omega\in\mathcal{W}}\Big|\frac{1}{V}\sum_{v=1}^V\Big[\mathcal{P}_{y_{v}}(\omega)\Big]  - \frac{1}{K}\sum_{k=1}^K\Big[\mathcal{P}_{\mathcal{H}_{\theta_{k}}[s_k](t) + \epsilon_k(t)}(\omega)\Big]\Big|,
\end{aligned}
\end{equation}
where $(\theta_k,s_k,\epsilon_k)\sim p(\theta, s, \epsilon | \nu)$, for $k=1,...,K$. Clearly, \eqref{eqn:sim_based_moment_matching} is not differentiable in $\nu$, so we use a zero-th order optimizer to estimate the minimizer \citep{frazier2018tutorial}.

\subsection{Uncertainty Quantification}\label{ssec:uq}
Let $\widehat{\tilde{\boldsymbol{\theta}}}$ denote the minimizer of \eqref{eqn:negLogPost}, and define $\widehat{\boldsymbol{\theta}} = g^{-1}(\widehat{\tilde{\boldsymbol{\theta}}})$. Although not the primary focus of this work, our goal is to provide a tool to quantify the uncertainty of this estimate at a prespecified location $x_v$, that is, to construct intervals for $\theta_j(x_v)$.
\par 
Under the proposed Bayesian formulation, the full posterior \eqref{eqn:posteriorlearnedSummary} provides a natural approach to uncertainty quantification. However, its high-dimensionality and non-convexity make MCMC-based approaches to sampling \eqref{eqn:posteriorlearnedSummary} computationally challenging. Moreover, the resulting credibility intervals would not necessarily be well calibrated in the frequentist sense, \revised{due to bias from the prior regularization}. Perhaps unsurprisingly, we found common approximations, such as the Laplace approximation, to substantially underestimate uncertainty.
\par 
We therefore propose a double bootstrap based approach, adapting the method from \cite{hall2013simple}, originally developed for nonparametric regression, to our current setting. \revised{Our adaptation resamples the observed signals using the stationary block bootstrap \citep{politis1994stationary}. Unlike the classical bootstrap for independent data, this method resamples temporally contiguous blocks of observations to approximately preserve the time dependence structure of the signals. An outer bootstrap constructs candidate pointwise $1-\alpha_0$  confidence intervals for the estimator at each $x_v$. For each outer resample, the inner bootstrap constructs a corresponding interval, and the proportion of these intervals that contain the original estimate is used to estimate the true coverage of the candidate interval. A calibrated level $\widehat{\alpha}$ is selected such that the bootstrap estimated coverage equals the desired level $1-\alpha_0$. This calibration step typically widens the intervals, correcting for undercoverage due to bias from the prior regularizer.} The algorithm is outlined below.

\begin{enumerate}
    \item Generate $B$ resamplings of the observed signals $Y^{(b)}$, $b=1,...,B$ using the stationary bootstrap. 
    \item For each bootstrapped sample $b=1,...,B$ 
    \begin{enumerate}
        \item Estimate $\widehat{\tilde{\boldsymbol{\theta}}}^{(b)}$ by minimizing \eqref{eqn:negLogPost} conditioned on $Y^{(b)}$
        \item For $r=1,...,R$, generate inner block-bootstrap resamples $Y^{(b,r)}$ from $Y^{(b)}$ using the stationary bootstrap. For each inner resample, estimate $\widehat{\tilde{\boldsymbol{\theta}}}^{(b,r)}$
        by minimizing \eqref{eqn:negLogPost} conditioned on $Y^{(b,r)}$.
    \end{enumerate}
    \item For each $j$, calculate the marginal variance at location $x_v$, $s_{\tilde{\theta}_{vj}}^2$, from the outer bootstrap estimates $\{\widehat{\tilde{\boldsymbol{\theta}}}_v^{(b)}\}_{b=1}^B$, and the variance $s_{\tilde{\theta}_{vj}^{(b)}}^2$ from the inner bootstrap estimates $\{\widehat{\tilde{\boldsymbol{\theta}}}_v^{(b,r)}\}_{r=1}^R$.
    \item For each $j$, estimate the empirical coverage at location \(x_v\) using the bootstrap coverage 
    $$
    \begin{aligned}
    \pi_{B}^{(j)}(x_v,\alpha) = \frac{1}{B}\sum_{b=1}^B\mathbb{I} \Big\{\widehat{\tilde{\boldsymbol{\theta}}}_{vj}\in &\Big(\widehat{\tilde{\boldsymbol{\theta}}}_{vj}^{(b)} - z_{1-\frac{\alpha}{2}}s_{\tilde{\theta}_{vj}^{(b)}},\widehat{\tilde{\boldsymbol{\theta}}}_{vj}^{(b)} + z_{1-\frac{\alpha}{2}}s_{\tilde{\theta}_{vj}^{(b)}}\Big) \Big\}
    \end{aligned}
    $$
    \item For target coverage $\revised{1-}\alpha_0$, determine $\widehat{\alpha}^{(j)}(x_v,\alpha_0)$ such that $\pi_{B}^{(j)}(x_v,\alpha) \approx 1-\alpha_0$. In practice, this is achieved by computing $\pi_{B}^{(j)}(x_v,\alpha)$ on a dense grid of candidate $\alpha$'s. We then define $\widehat{\alpha}_{\xi}^{(j)}(\alpha_0)$ to be the $\xi$-level quantile of $\{\widehat{\alpha}^{(j)}(x_v,\alpha_0):x_v\in\mathcal{X}\}$. 
    In practice, we set \revised{$\xi=0.025$}. 
    \item Return the transformed bootstrap calibrated pointwise intervals 
    $$
    \begin{aligned}
        \Big\{ \Big(&g_j^{-1}\Big(\widehat{\tilde{\boldsymbol{\theta}}}_{vj} - z_{1-\frac{\widehat{\alpha}^{(j)}_{\xi}(\alpha_0)}{2}}s_{\tilde{\theta}_{vj}}\Big), g_j^{-1}\Big(\widehat{\tilde{\boldsymbol{\theta}}}_{vj} + z_{1-\frac{\widehat{\alpha}^{(j)}_{\xi}(\alpha_0)}{2}}s_{\tilde{\theta}_{vj}}\Big)\Big):x_{v}\in\mathcal{X} 
        \Big\}.
    \end{aligned}
    $$
\end{enumerate}

\subsection{\revised{Signal Deconvolution}}\label{ssec:signal_deconv}
\revised{The proposed methodology is primarily focused on estimation of the cortical hemodynamic field. Neural signal deconvolution, if desired, can then be performed conditional on this estimate by constructing the corresponding spatially varying forward model and applying an off-the-shelf non-blind deconvolution method, for which a variety of approaches are available \citep{Uru_uela_2023}. This modularity allows the hemodynamic estimation procedure to be paired flexibly with different downstream deconvolution methods.}
\par 
\revised{In practice, however, it may be useful to provide the user a default deconvolution procedure. Since the non-blind inverse problem is still highly ill-posed, regularization is required to define a stable estimator. To align the regularization with the sparse, ``spike-like'' neural signal model in Section~\ref{ssec:model_neural_signals}, 
we use an $\ell_1$-penalized regression approach \citep{caballero2013paradigm}, applied independently at each mesh vertex under the estimated forward model. The resulting optimization problems are solved using the FISTA algorithm \citep{beck2009fast} implementation from the \textit{pySPFM} package. The regularization parameter is selected using the package's default percentage-of-maximum criterion, which is set to $80\%$ of the smallest parameter that results in the zero solution \citep{friedman2010regularization}. 
}
\section{Experiments and Implementation}\label{sec:experiments}

\subsection{Hemodynamic Models}

\begin{figure}[t]
  \centering
  \begin{minipage}[t]{0.49\linewidth}
    \centering
    \includegraphics[width=\linewidth]{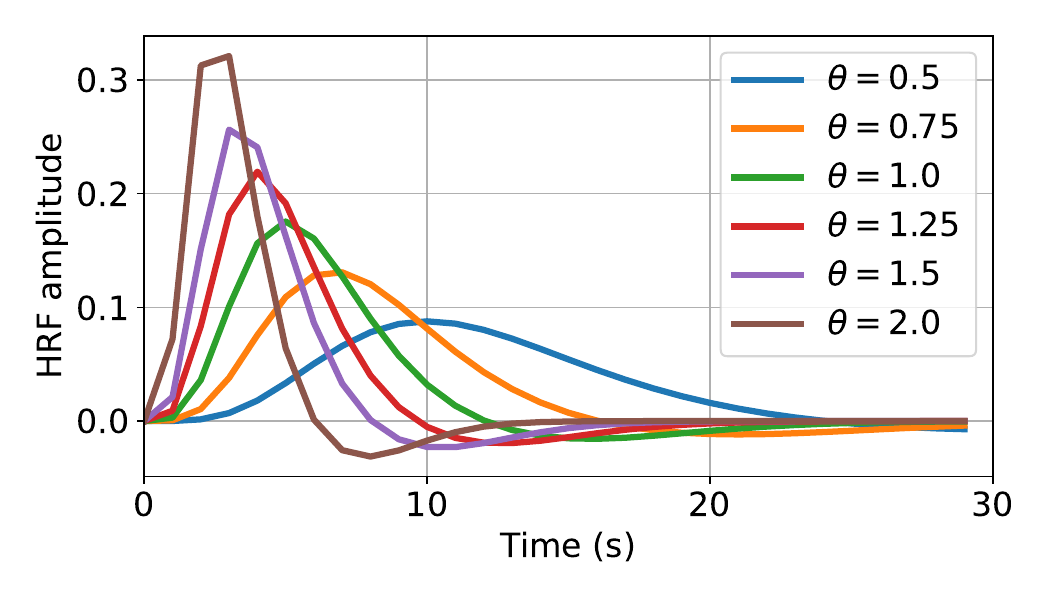} 
  \end{minipage}\hfill
  \begin{minipage}[t]{0.49\linewidth}
    \centering
    \includegraphics[width=\linewidth]{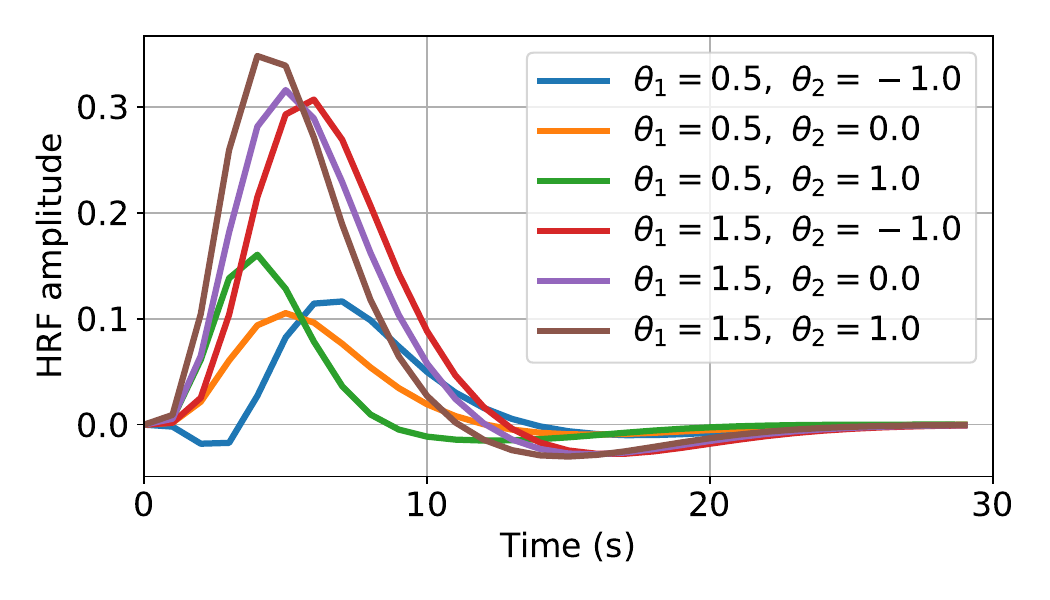} 
  \end{minipage}
  \caption{Hemodynamic response kernel for several $\theta$ for model \eqref{eqn:double_gamma_hrf} (left panel) and \eqref{eqn:canonical_gamma_basis} (right panel).}
  \label{fig:hrf_1param}
\end{figure}

To model the hemodynamic response, we assume an LTI forward model of the form \eqref{eqn:LTI_HR_model} under two popular parametric kernels built from the so-called canonical double Gamma function, defined as  
\begin{equation}\label{eqn:canonical_HRF}
    h(t) = \frac{t^{a_1-1}b_1^{a_1}\exp(-b_1 t)}{\Gamma (a_1)} - c \frac{t^{a_2-1}b_{2}^{a_{2}}\exp(-b_2 t)}{\Gamma (a_2)}
\end{equation}
with constants $a_1=6$, $a_2=16$, $b_1=1$, $b_2=1$, $c=1/6$. The first model is a one-parameter shifted double gamma model \citep{Cherkaoui2021}, given by
\begin{equation}\label{eqn:double_gamma_hrf}
    \revised{h_{\theta}(t) = \frac{\theta^{a_1}t^{a_1-1}\exp(-\theta t)}{\Gamma (a_1)} - c \frac{\theta^{a_2}t^{a_2-1}\exp(-\theta t)}{\Gamma (a_2)}},
\end{equation}
with parameter bounds $\theta\in [0.5, 2.5]$. The second is a two-parameter \revised{model} of the form 
\begin{equation}\label{eqn:canonical_gamma_basis}
        h_{\theta}(t) = \theta_1h(t) + \theta_2\frac{\partial}{\partial t}h(t),
\end{equation}
for which reasonable physiological ranges are $\theta_1\in [0.2, 2.0]$, $\theta_2\in [-1.0, 1.0]$ \citep{lindquist2009modeling}.
\par 
\revised{Both models are low-dimensional perturbations of the canonical double Gamma function~\eqref{eqn:canonical_HRF}. The one-parameter model in \eqref{eqn:double_gamma_hrf} is obtained from~\eqref{eqn:canonical_HRF} by replacing the fixed rate parameters $b_1=b_2=\theta$. The two parameter model \eqref{eqn:canonical_gamma_basis} assumes the HRF is in the linear span of a pair of basis functions, \eqref{eqn:canonical_HRF} and its time derivative.} Figure~\ref{fig:hrf_1param} shows how the kernel shape changes with respect to $\theta$ for both models. 

\subsection{Datasets}

\begin{figure}[!ht]
    \centering
    \includegraphics[width=\textwidth]{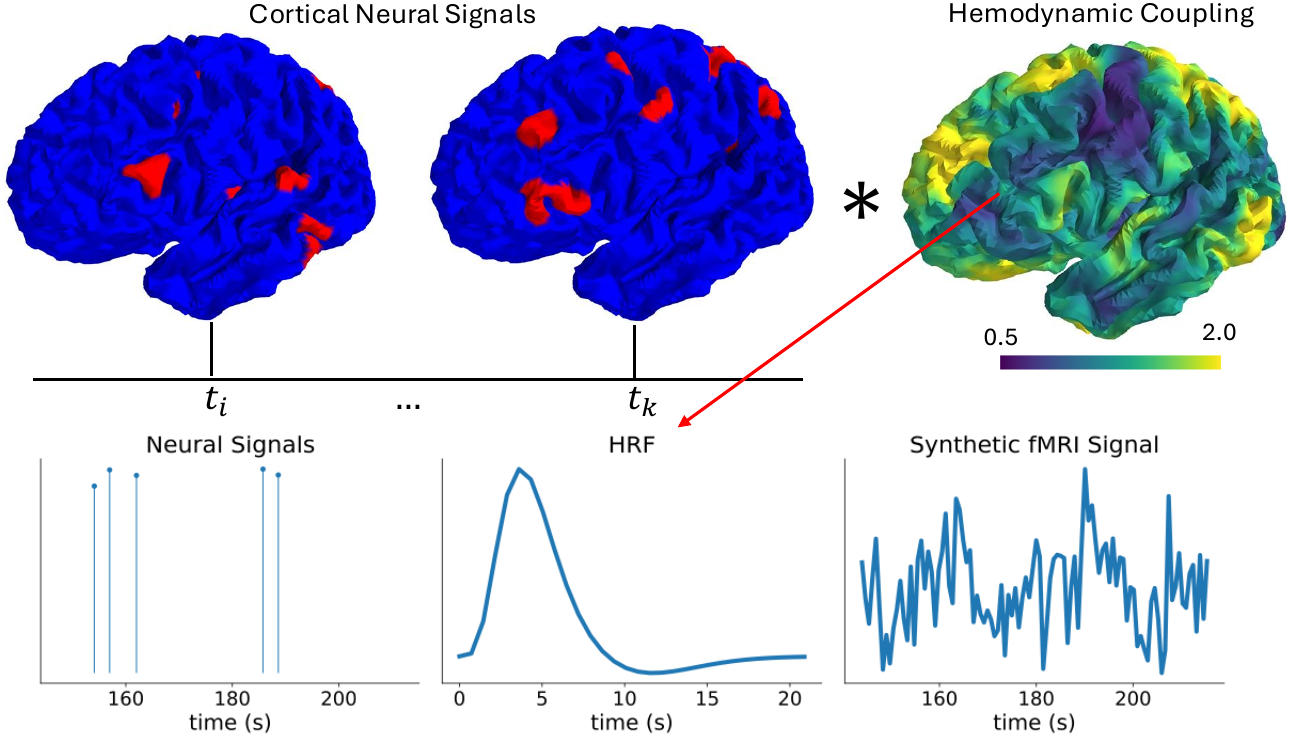}
    \caption{\revised{Sample of the synthetic data. (Top left) Spatially correlated sparse cortical neural activity (red regions) at two sample time slices. (Top right) Sample hemodynamic coupling parameter field for HRF model \eqref{eqn:double_gamma_hrf}. (Bottom) Sample neural signals, HRF kernel and corresponding synthetic noisy BOLD signals at the indicated mesh vertex.}}
    \label{fig:synethic_data_generation}
\end{figure}

\subsubsection{In-vivo Data}
We use data from the publicly available Young Adult Human Connectome Project (HCP-YA) \citep{glasser2016human} to evaluate our method. Specifically, we use the preprocessed rsfMRI data, denoised with ICA-FIX
 \citep{salimi2014automatic}, and mapped to the left cortical surface mesh with $32,492$ vertices and registered using FreeSurfer. Details on the acquisition and preprocessing protocols can be found in the standard reference \citep{glasser2013minimal}, and are thus omitted here. 

\subsubsection{Synthetic Data}
\revised{\textbf{Neural Activity}: Cortical neural activity exhibits complex spatial dependence. To induce spatially correlated neural activity in synthetic data, we used the following procedure. At each time point, a random binary vector $z\in \{0,1\}^{V}$ was generated and spatially smoothed by solving the penalized least-squares problem
$$
\zeta^{*} = \text{argmin}_{\zeta\in\mathbb{R}^{V}}\|z - \zeta\|_{2}^2 + \rho \zeta^{\intercal}G^{\intercal}C^{-1}G\zeta,
$$
where $C,G$ are the matrices defined in  \eqref{eqn:mass_stiffness_matrices}, and $\rho>0$ determines the amount of smoothing. The penalty term is a discrete approximation to the second order roughness penalty based on the Laplace-Beltrami operator \citep{Lila2016}. Vertices for which $\zeta^{*}$ exceeded a small threshold  were used to identify the non-zero regions of neural activity, and neural signal amplitudes at these vertices were simulated using the calibrated amplitude ranges. Figure~\ref{fig:synethic_data_generation} (top left) shows representative spatial patterns of the simulated neural signals at a couple time points.}
\par 
\revised{\textbf{LTI Hemodynamics}: We simulate hemodynamic responses using both the 1-parameter \eqref{eqn:double_gamma_hrf} and 2-parameter~\eqref{eqn:canonical_gamma_basis} HRF kernels. For the 1-parameter model, synthetic hemodynamic parameter fields $\tilde{\theta}$ were simulated using the GP prior outlined in Section~\ref{ssec:spatial_prior}, with $\kappa=8\cdot10^{-2}$, $\tau^2=10^{2}$. For the 2-parameter model~\eqref{eqn:canonical_gamma_basis}, we use a GP with $\kappa_1=\kappa_2=8\cdot10^{-2}$, $\tau_1^2=\tau^2_2=4\cdot 10^1$ and add a random, spatially constant offset sampled from a tightly peaked normal distribution around $(-0.67, 0.32)$.}
\par 
\revised{\textbf{Nonlinear Hemodynamics}: To evaluate performance when the assumed LTI hemodynamic forward model is misspecified, we also simulate BOLD responses using the Balloon-Windkessel  model, a nonlinear coupled system of ODEs linking neural activity to the BOLD signal \citep{friston2000nonlinear}. To simulate spatially dependent hemodynamics, we allow two parameters to vary over the cortex, the vasodilatory signal decay and flow-dependent feedback rate, as previous studies have found these parameters to be influential for determining the hemodynamic response in task-based fMRI \citep{zayane2015sensitivity}. The remaining model parameters are fixed at nominal values found in the literature \citep{ziaeemehr2025virtual}. The parameter field was simulated on the surface using a GP in the transformed space with smoothness $\kappa_1=\kappa_2=8\cdot10^{-2}$, $\tau^2_1=\tau^2_2=4\cdot10^{1}$. The ODE forward model was integrated using the Euler method with timestep $dt=0.05$(s). Since neural events were simulated on the much coarser acquisition grid ($t_r = 0.72$(s)), each event was held 
constant over the corresponding sub-interval during integration.  
}
\par 
\revised{\textbf{rsfMRI Signals}: The calibration procedure from Section~\ref{sssec:simulator_calibration} was run on the rsfMRI data of a randomly selected HCP-YA subject under both LTI hemodynamic forward models, and the results were used to parameterize the measurement error model (Section~\ref{ssec:statistical_model}). For the Balloon-Windkessel model, the measurement error variance was set so that the resulting mean SNR matched the mean SNR obtained under the calibrated noise level for HRF model~\eqref{eqn:canonical_gamma_basis}. 
The final synthetic rsfMRI time series were generated at all $V=32,492$ left cortical surface mesh vertices for $M=1200$ time points with $t_r=0.72$s, matching the HCP-YA protocol.}
\subsection{Implementation details}\label{ssec:implementation_details}
For the summary network $T_{\psi}(\cdot)$, we use a fully connected network with ReLU activations, with an initial non-trainable encoding layer which takes the Fourier transform of the BOLD signals $y$ at the set of Nyquist frequencies\revised{, given in \eqref{eqn:nyquist_frequencies}}. We set the number of hidden layers to be 3, with hidden widths $M,\lfloor \frac{M}{2}\rfloor, \lfloor \frac{M}{4}\rfloor$. For the marginal likelihood emulator $p_{\gamma}(\cdot|\cdot)$, we use a conditional neural spline normalizing flow \citep{durkan2019neural} with $5$ spline coupling transforms, each with a fully connected 3 layer conditioning network with hidden widths of 64 and ReLU activations. Both networks are trained on synthetic data generated from the simulator outlined in Section~\ref{sec:models} using Adam with a batch size of $100$ and learning rate of $10^{-5}$ for $10^{5}$ iterations. Note that separate summary network and likelihood model pairs were trained for the forward models \eqref{eqn:double_gamma_hrf} and \eqref{eqn:canonical_gamma_basis}. 
\par 
For calibrating the simulator hyperparameters using the non-differentiable loss \eqref{eqn:sim_based_moment_matching}, we used Bayesian optimization \citep{frazier2018tutorial} under the expected improvement acquisition function for a total of $200$ runs using the rsfMRI from a randomly selected HCP-YA subject and both forward models. The prior smoothness hyperparameters are selected by maximizing \eqref{eqn:margLik}
over a pre-defined grid. For single subject analysis, this was repeated for each subject. For multi-subject analysis, we use a common set of smoothing parameters for all reconstructions, as is standard practice. 

\subsection{Competing Hemodynamic Estimators}

For estimation under the one-parameter model~\eqref{eqn:double_gamma_hrf}, we compare our method against two alternative approaches. The first approach,  referred to as \textit{JointMAP}, performs vertex-wise joint MAP reconstruction under a standard $l_1$ sparsity prior on the signals
$$
\left(\widehat{\theta}_v, \widehat{s}_v \right) = \text{min}_{\theta \in\Theta,s\ge 0} \left\| y_v - A(\theta)s \right\|^2 + \eta \|s\|_1,
$$
where $A(\theta) \in \mathbb{R}^{M\times M}$ is a convolution matrix formed from the kernel \eqref{eqn:double_gamma_hrf}, $s\in\mathbb{R}^{M}$ denotes the discretized neural signals, and $\eta >0$ is a regularization parameter. Optimization is performed using  block-coordinate descent, where the (convex) subproblem in $s$ is solved using alternating direction method of multipliers and the (non-convex) subproblem in $\theta$ is solved using golden section search. We set the penalty parameter $\eta=\sigma\sqrt{2\log(M)}$, where $\sigma$ is the calibrated noise level. The second approach, referred to as \textit{DUP} (deep unrolled prior), adapts  
\cite{Gossard2024} to perform vertex-wise joint reconstruction. Specifically, DUP first pre-trains an unrolled network $\mathcal{A}_{\rho}:\mathbb{R}^{M}\times\mathbb{R}^{M\times M}\mapsto\mathbb{R}^{M}$, parameterized by weights $\rho$, to minimize
$$
    \widehat{\rho} = \min_{\rho}\mathbb{E}_{p(\theta,s,y)}\left\|\mathcal{A}_{\rho}(y, A(\theta)) - s \right\|_2^2,
$$
then estimates the hemodynamic parameters by minimizing the reconstructed error 
\begin{equation}\label{eqn:DUP_inference}
    \widehat{\theta}_v = \min_{\theta\in\Theta}\|A(\theta)\mathcal{A}_{\widehat{\rho}}(y_v, A(\theta)) - y_v\|_{2}^2, 
\end{equation}
via golden section search. We take $\mathcal{A}_{\rho}$ to be a 5-layer unrolled fully connected ReLU network with hidden width $M$ and train on synthetic data from the simulator model in Section~\ref{sec:models} using Adam (learning rate $10^{-5}$, $10^{5}$ iterations, batch size $100$). 
We also attempted to compare our method against \cite{Cherkaoui2021}. However, extending their method to a high-resolution surface mesh proved computationally intractable, which is not surprising as the approach was originally developed to work on coarse, atlas-based surface discretizations.
\par 
For the two-parameter model~\eqref{eqn:canonical_gamma_basis}, we compare our method to \cite{WU2013365}, implemented in the MATLAB package \textit{rsHRF} \citep{WU2021118591}. rsHRF uses a two stage procedure, first estimating neural spike times by thresholding the fMRI time series, then estimating hemodynamic parameters conditional on the detected spikes. We used the suggested defaults for all method hyperparameters. 

\subsection{Evaluation}

\subsubsection{\revised{Hemodynamic Coupling Inference}}

\revised{We use synthetic datasets to evaluate the performance of the competing hemodynamic estimation methodologies in both correctly and misspecified settings. In the correctly specified setting, where the parametric form of the hemodynamic response is the same for fMRI signal generation and model estimation, we compute the relative mean squared error (rMSE) and bias of the estimated HRF parameters, defined by 
$$
rMSE(\widehat{\boldsymbol{\theta}}_j) = \frac{\|\widehat{\boldsymbol{\theta}}_j - \boldsymbol{\theta}_j\|_{2}^2}{\|\boldsymbol{\theta}_j\|_{2}^2}, \quad \text{Bias}(\widehat{\boldsymbol{\theta}}_j) = \frac{1}{V}\sum_{v=1}^V(\widehat{\boldsymbol{\theta}}_{j,v}-\boldsymbol{\theta}_{j,v}), 
$$
respectively.
The quality of the uncertainty quantification procedure from Section~\ref{ssec:uq} is assessed by the empirical coverage of the $95\%$ pointwise intervals\revised{, which is} computed as the fraction of mesh locations for which the intervals cover the true hemodynamic parameter. We also compute the average interval length over the mesh for use as a scalar measure of reconstruction uncertainty.}
\par
\revised{To evaluate the misspecified setting, we use our parametric kernels to fit the synthetic fMRI signals generated by the Balloon-Windkessel model. In this setting, the fitted and true hemodynamic responses no longer share the same parameter space, so we quantify estimation performance using the estimated HRF kernel itself. Since the Balloon-Windkessel 
model is nonlinear, it does not admit a single HRF, so instead we approximate the response at each vertex by integrating the forward model for an isolated neural event. The left panel of Figure~\ref{fig:synth_misspec} shows several of the resulting approximate kernels corresponding to different parameters. To allow fair comparison across HRF parameterizations and remove scale ambiguity between the HRF and neural-signal amplitude under the LTI model, we use the average correlation dissimilarity, defined as,
\begin{equation}\label{eqn:hrf_Pearson_correlation}
  \text{d}_{\text{HRF}}(h^{(1)}, h^{(2)}) = {V}^{-1}\sum_{v=1}^V\left[1 - \text{corr}\left(h_{v}^{(1)},h_{v}^{(2)}
  \right)\right],
\end{equation}
where $h^{(1)},h^{(2)}$ are HRF kernels at vertex $x_v$, with potentially different parametric forms, and $\text{corr}$ is the Pearson correlation.}
\par
\revised{Each synthetic experiment is repeated for $10$ Monte-Carlo replications, and averaged results are reported.}

\subsubsection{\revised{Reproducibility Analysis}}

\revised{For in-vivo data, where we lack access to ground truth, we evaluate the test-retest reproducibility of our methodology. Since the hemodynamic response shape is expected to be relatively stable within healthy subjects between multiple scans over a short time period (months), the HCP test-retest scans can be (approximately) considered as independent replications from model \eqref{eqn:bold_forward_model} with the same $\theta$, though different $s$ and $\epsilon$, and thus can be used to assess the stability of our hemodynamic coupling estimation methodology. To quantify reliability, we use the dbICC metric \citep{xu2021generalized}, defined by, 
\begin{equation}\label{eqn:dbICC}
\text{dbICC} = 1 - \frac{\text{MSD}_{\text{within}}}{\text{MSD}_{\text{between}}},
\end{equation}
where $\text{MSD}_{\text{between}}$, $\text{MSD}_{\text{within}}$ denote the mean between- and within- subject squared dissimilarities, respectively, computed using \eqref{eqn:hrf_Pearson_correlation}.}
\par 
\revised{We additionally assess the stability of downstream hemodynamic inversion using seed-based functional connectivity (FC) estimated from the deconvolved neural signals, reconstructed using the default approach in Section~\ref{ssec:signal_deconv}. Specifically, for each left-hemisphere seed region in the Glasser atlas \citep{glasser2016multi}, the Pearson correlation between the seed and each non-seed surface mesh vertex is computed. The reproducibility is quantified using the dbICC~\eqref{eqn:dbICC} with $l_2$ distance on the FCs over the cortical surface.}

\subsubsection{Population Analysis}

We evaluate inter-subject variation of the hemodynamic field estimates using the functional principal components analysis (fPCA) approach proposed in \citep{Lila2016}.
\revised{As displayed in Figure~\ref{fig:hrf_1param}, $\theta$ controls key shape characteristics of the HRF, particularly its latency, that is, the time delay of the BOLD response following a neural event. For interpretability, we transform the parameter fields into spatial latency fields by computing the pointwise \textit{time-to-peak}: $l(x) := \text{argmax}_{t}h_{\theta(x)}(t)$.} For a cohort of $N$ subjects, fPCA is used to obtain the Karhunen–Loève decomposition of the \revised{time-to-peak $l_i(x_v)$} for subject $i$ at location $x_v$:
\begin{equation}\label{eqn:latent_KL}
    \revised{l_{i} (x_v) = \mu_{l}(x_v) + \sum_{k=1}^{N-1} z_{ik}\varphi_k(x_v)},
\end{equation}
where \revised{$\mu_l(x_v)$} is the mean function, \revised{$\varphi_k(x_v)$} are orthonormal eigenfunctions, and the scores $z_{ik}$ are mean-zero, uncorrelated random variables with $\mathrm{Var}(z_{ik}) = \lambda_k$. The eigenvalues ${\lambda_k}$ are non-increasing in $k$. The first few eigenfunctions represent the dominant modes of spatial variation in the \revised{latency} across subjects, while the corresponding scores quantify the contribution of each mode for an individual subject. 
\par
We use the scan-1 rsfMRI data from $N=100$ unrelated HCP-YA subjects to compute \revised{$\{{\varphi}_k\}_{k=1}^{N-1}$} and the corresponding subject specific score $\{{z}_{ik}\}_{k=1,i=1}^{N-1,N}$. fPCA implies a reduced rank estimate of the covariance of \revised{$l(x)$}, which is given by 
\begin{equation}\label{eqn:low_rank_cov}
    \revised{\text{Cov}_{l}(x_j,x_{j'}) \approx \sum_{k=1}^{N-1}{\lambda}_k{\varphi}_k(x_j){\varphi}_k(x_{j'}), \qquad j,j' \in 1,\ldots, V.}
\end{equation}
This will be used to identify cortical regions with large inter-subject hemodynamic variability.  

\section{Results}\label{sec:results}

\subsection{Synthetic Data Results}\label{ssec:synth_results}

\begin{table}[t]
\centering
\begin{tabular}{@{}lcccc@{}}
\toprule
& JointMAP & DUP & MPM & Ours \\
\midrule
\revised{rMSE}  & \revised{0.4772 (0.0415)} & \revised{0.4853 (0.0404)} & \revised{0.0405 (0.0012)} & \revised{0.0068 (0.0001)} \\
Bias & \revised{0.9914 (0.0298)} & \revised{0.9200 (0.0316)} & \revised{-0.0170 (0.0072)} & \revised{-0.0372 (0.0056)} \\
\bottomrule
\end{tabular}
\caption{\revised{Synthetic-data results for the one-parameter model \eqref{eqn:double_gamma_hrf}. Results averaged across cortical surface and over Monte Carlo repetitions. Standard errors reported in parentheses.}}
\label{tab:synth_1}
\end{table}

\begin{table}[t]
\centering
\begin{tabular}{@{}llcc@{}}
\toprule
\revised{Parameter} & Metric & rsHRF & Ours \\
\midrule
\multirow{2}{*}{$\theta_1$}
    & \revised{rMSE}  & \revised{0.9009 (0.0013)}  & \revised{0.0124 (0.0003)} \\
    & Bias & \revised{-0.7581 (0.0073)} & \revised{-0.0280 (0.0014)} \\
\addlinespace
\multirow{2}{*}{$\theta_2$}
    & \revised{rMSE}  & \revised{0.9661 (0.0023)}  & \revised{1.2652 (0.0378)} \\
    & Bias & \revised{0.1779 (0.0163)}  & \revised{0.1318 (0.0182)} \\
\bottomrule
\end{tabular}
\caption{\revised{Synthetic-data results for the two-parameter model \eqref{eqn:canonical_gamma_basis}. Results averaged across cortical surface and over Monte Carlo repetitions. Standard errors reported in parentheses.}}
\label{tab:synth_2}
\end{table}

\begin{table}[t]
\centering
\begin{tabular}{@{}lccc@{}}
\toprule
& \multicolumn{1}{c}{Model \eqref{eqn:double_gamma_hrf}}
& \multicolumn{2}{c}{Model \eqref{eqn:canonical_gamma_basis}} \\
\cmidrule(lr){2-2}\cmidrule(lr){3-4}
& $\theta_1$ & $\theta_1$ & $\theta_2$ \\
\midrule
EC              & \revised{0.9499 (0.0056)} & \revised{0.9186 (0.0042)}  & \revised{0.9536 (0.0053)} \\
Interval Length & \revised{0.5488 (0.0043)} & \revised{0.3085 (0.0050)} & \revised{1.5294 (0.0201)} \\
\bottomrule
\end{tabular}
\caption{\revised{Uncertainty quantification results. Point-wise empirical coverage (EC) and average interval length averaged Monte Carlo experimental repetitions. Standard errors are reported in parentheses.}}
\label{tab:synth_uq}
\end{table}


\begin{figure}
    \centering
    \includegraphics[width=\textwidth]{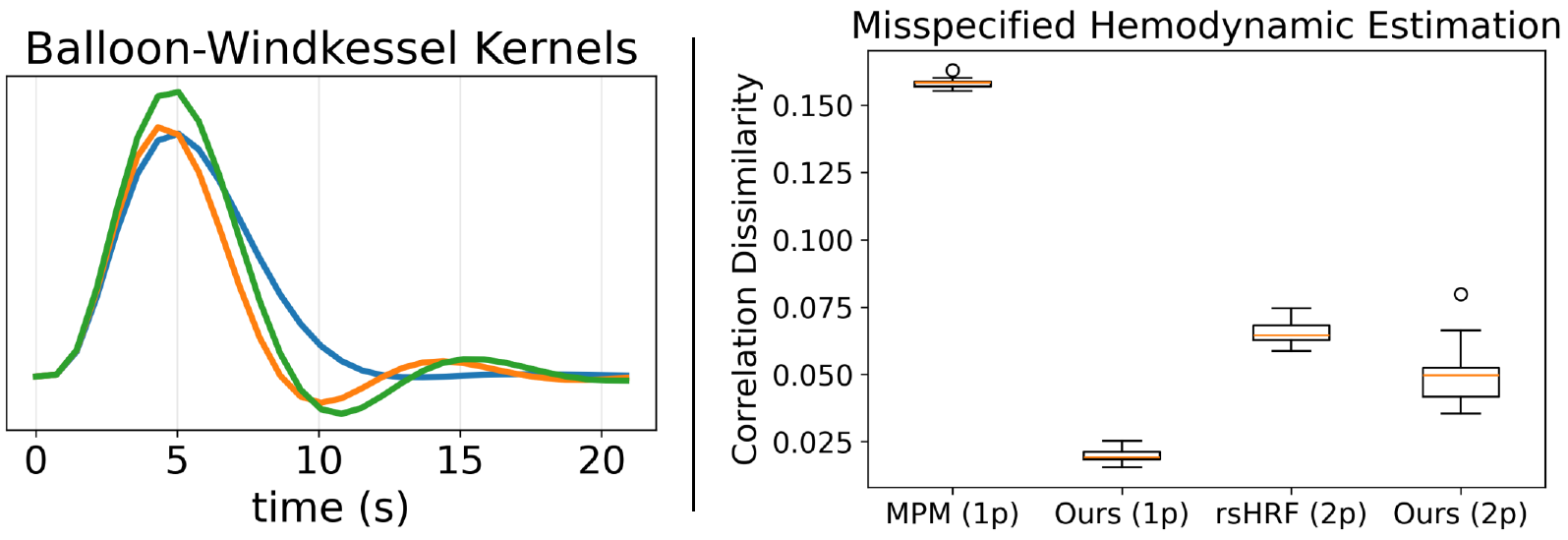}
    \caption{\revised{(Left) Approximate  Balloon-Windkessel hemodynamic responses. Each color is a different parameter combination. (Right) Synthetic-data results for the misspecified hemodynamic setting. Boxplots of Monte Carlo correlation dissimilarity $d_{HRF}$ between the approximate Balloon-Windkessel kernel and the HRF estimated by different methods.}}
    \label{fig:synth_misspec}
\end{figure}

\revised{Table~\ref{tab:synth_1}} compares the estimation results for the synthetic data using the one-parameter model \eqref{eqn:double_gamma_hrf}. Our method exhibits \revised{rMSEs} and biases that are, on average, an order of magnitude smaller than both JointMAP and DUP. Recall that both of the competitors perform joint reconstruction of signal and hemodynamic parameters, resulting in the estimation of the latter being dependent on the estimation of the former. Since even the non-blind signal reconstruction is ill-posed in this setting, we attribute the poorer performance of the competing methods, at least in part, to the propagation of errors from the signal reconstruction to the target hemodynamic estimation. Our approach avoids this issue by marginalizing the signal out of the data likelihood. \revised{Recall that for this marginalization to be computationally tractable, we used an approximation of spatially uncorrelated neural signals. This assumption is violated by our data generative model, which synthesizes neural signals with substantial spatial correlation. The strong estimation performance demonstrated by our method indicates robustness to significant departures from the invoked independence approximation, which is important in practice as cortical neural activity is likely strongly spatially correlated.}
\par 
The second key differentiating feature of our method is the incorporation of spatial regularization. To evaluate its effect, we drop the prior from Section~\ref{ssec:spatial_prior} and apply the learned marginal posterior mean \eqref{eqn:posterior_mean_summary}, denoted MPM, independently to each mesh BOLD time series. As shown in the third row of \revised{Table~\ref{tab:synth_1}}, on average, this results in a \revised{roughly} $6\times$ increase in the \revised{rMSE} compared to our spatially regularized framework. While MPM outperforms both JointMAP and DUP owing to its avoidance of explicit signal reconstruction, we see that incorporating spatial smoothness of the hemodynamics provides substantial additional improvement. \revised{Note that this improvement in rMSE is at the expense of a modest increase in bias. This is not unexpected, as incorporating spatial regularization can results in some bias in the parameter estimate. However, such regularization can also substantially reduce the variance of the estimator, leading to an overall reduction in MSE, which is what is observed in our results.}
\par 
\revised{Table~\ref{tab:synth_2}} shows the estimation results for the two-parameter forward model \eqref{eqn:canonical_gamma_basis}. Estimation of $\theta_1$ echoes the results from \revised{Table~\ref{tab:synth_1}}, where our method again results in an order of magnitude reduction in average \revised{rMSE} compared to rsHRF. \revised{rsHRF exhibits lower rMSE for $\theta_2$, though both approaches perform quite poorly for this parameter. Indeed, both are roughly on par with the trivial estimator of simply predicting the parameter prior mean at each mesh vertex, i.e. $\widehat{\theta}_2(x)=0.0$, which results in an rMSE of $1$.} 
\par 
\revised{Table~\ref{tab:synth_uq} C) shows the $95\%$ empirical pointwise coverage and average interval length for the procedure from Section~\ref{ssec:uq} for both synthetic datasets. For the one-parameter model~\eqref{eqn:double_gamma_hrf}, we see good empirical coverage. For the two-parameter model~\eqref{eqn:canonical_gamma_basis}, the coverage on $\theta_1$ indicates slight overconfidence, though the intervals are quite tight, indicating strong statistical identifiability of this parameter. While the empirical coverage for $\theta_2$ is good, the estimation uncertainty is excessively large. Indeed, given $\theta_2\in[-1,1]$, the average interval length occupies approximately $75\%$ of the possible values of the parameter. Coupled with the poor estimation performance for both methods for $\theta_2$, these results suggest potential statistical identifiability issues for this parameter at the considered sampling rate and noise levels. That is, there may not be enough information in the observed resting state signals under HCP-style acquisition and noise level to infer this parameter with reasonable precision.}
\par 
\revised{The right panel of Figure~\ref{fig:synth_misspec} shows the estimation results for several of the considered methods under hemodynamic forward-model misspecification. We see that the proposed approach under both considered hemodynamic kernels outperforms rsHRF. We also note that our method substantially outperforms MPM, further underscoring the importance of incorporating spatial regularization. These results provide evidence that the proposed framework is robust to some forms of hemodynamic model misspecification, particularly under the one-parameter HRF model~\eqref{eqn:double_gamma_hrf}. Robustness to such hemodynamic forward-model misspecification is important in practice, as some degree is likely unavoidable given the complexity of the underlying biophysics.}

\subsection{In-vivo Data Results}\label{ssec:invivo}

In this section, we apply our methodology under the one-parameter model~\eqref{eqn:double_gamma_hrf} to the real resting-state HCP data. We choose this model over model~\eqref{eqn:canonical_gamma_basis} due to \revised{the stronger statistical identifiability of its parameters and superior performance under hemodynamic forward model misspecification observed in the synthetic data experiments in Section~\ref{ssec:synth_results}.}

\subsubsection{\revised{Reproducibility Analysis}}

\begin{figure*}[!t] 
  \centering
  \includegraphics[width=\textwidth]{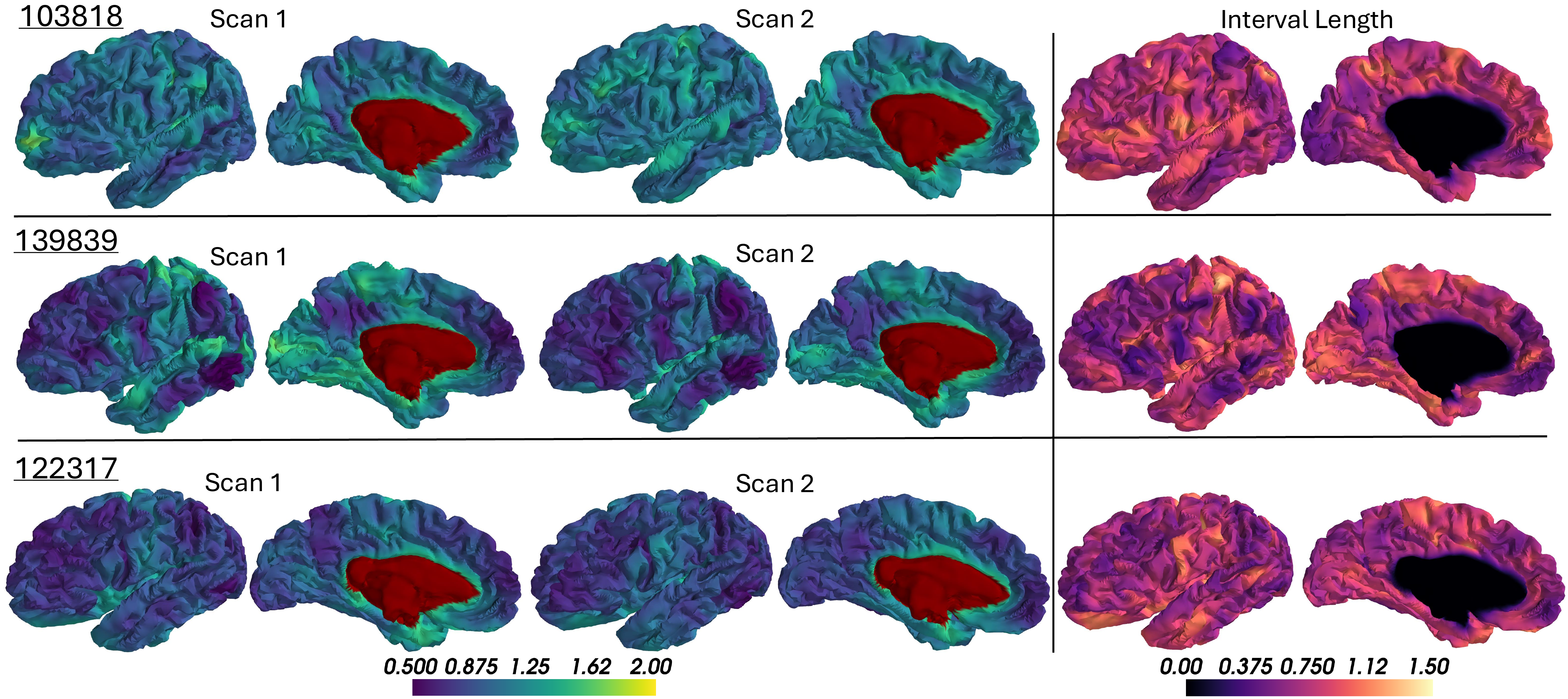}
  \caption{(Left) Estimates of the $\theta$-field for the one-parameter hemodynamic coupling model \eqref{eqn:double_gamma_hrf} on the HCP test-retest data for three randomly selected subjects. Inter-subject variability is noticeably larger than within-subject between-scan variability. (Right) Length of the $95\%$ pointwise intervals calculated from the procedure in Section~\ref{ssec:uq} on the Scan 1 data.}
  \label{fig:reproducibility}
\end{figure*}
\begin{figure}[t]
\centering

\begin{minipage}[c]{0.45\textwidth}
    \centering
    \captionsetup{width=0.95\linewidth}
    \vspace{0.5em}

    \begin{tabular}{lc}
    \toprule
    Method & HRF dbICC \\
    \midrule
    Ours  & $0.497\;(0.295,\;0.648)$ \\
    rsHRF & $0.140\;(0.0872,\;0.187)$ \\
    \bottomrule
    \end{tabular}
    \captionof{table}{\revised{HCP test-retest dbICC of the estimated
    hemodynamic response (95\% bootstrap confidence intervals in parentheses).}}
    \label{tab:hrf_reproducibility}
\end{minipage}
\hspace{0.03\textwidth}
\begin{minipage}[c]{0.45\textwidth}
    \centering

    \includegraphics[width=\linewidth]{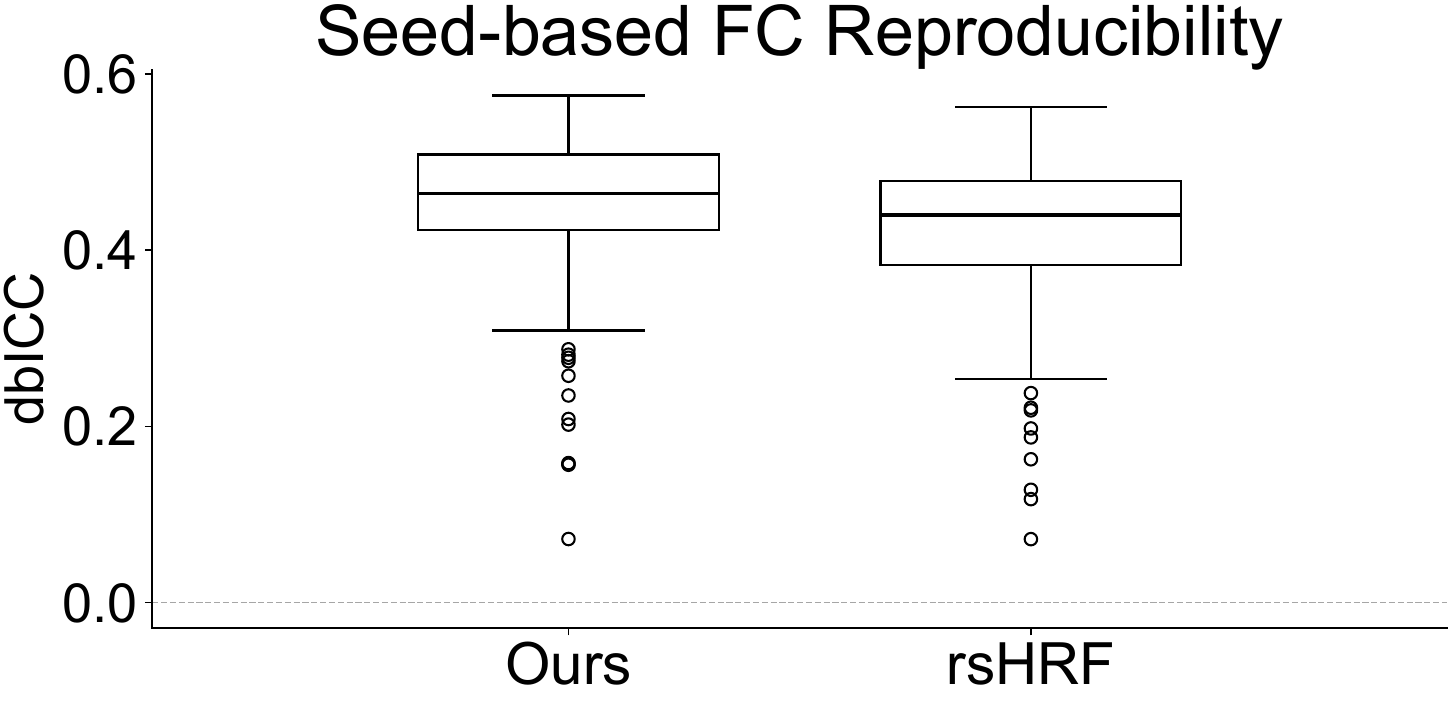}

    \captionsetup{width=0.95\linewidth}
    \captionof{figure}{\revised{Boxplot of HCP test-retest dbICCs of
    seed-based functional connectivity for all left-hemisphere parcels
    in the Glasser Atlas.}}
    \label{fig:seed_fc_dbICC}
\end{minipage}

\end{figure}

\revised{\textbf{Hemodynamic Coupling:}}
The spatial maps of $\widehat{\boldsymbol{\theta}}$ for 3 randomly selected HCP test-retest subjects are shown in the left part of Figure~\ref{fig:reproducibility}. Qualitatively, we see that between-subject variability is noticeably larger than within-subject (between-scan) variability for each subject, with many unique subject-specific spatial features preserved across scans. For each subject in Figure~\ref{fig:reproducibility}, we ran the uncertainty quantification procedure from Section~\ref{ssec:uq} on the scan 1 data to construct pointwise $95\%$ intervals for the reconstructed field. The resulting interval lengths are shown in the right part of Figure~\ref{fig:reproducibility}. We notice substantial spatial variability in these maps, indicating the reconstruction uncertainty is highly location dependent. \revised{Averaged over all spatial locations and subjects, $\approx 98.5\%$ of the scan 2 estimates are contained within the intervals constructed from scan 1. Since the true hemodynamic parameter is unknown, this quantity should not be interpreted as empirical coverage of the true parameter. However, its reasonably close agreement with the nominal level suggests that the intervals may provide a useful, albeit slightly conservative, measure of within-subject scan-rescan uncertainty. Table \eqref{tab:hrf_reproducibility} provides the HRF dbICC for our method and rsHRF estimated using all $45$ HCP test-retest subjects. The $95\%$ bootstrap confidence intervals are reported in parenthesis. Our method exhibits substantially greater test-retest reproducibility than rsHRF.}
 
\begin{figure}[!ht]
    \centering
    \includegraphics[width=\textwidth]{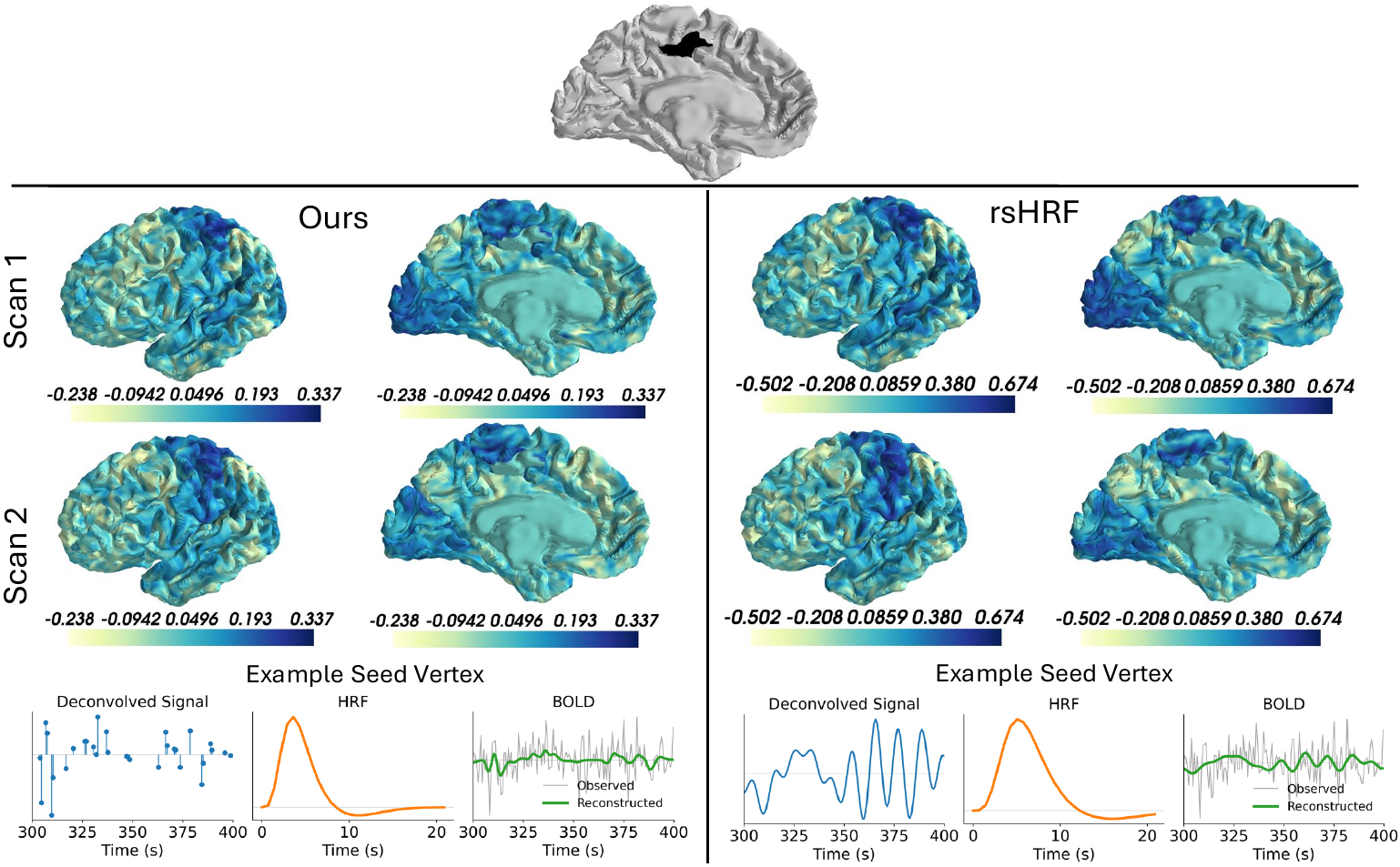}
    \caption{\revised{(Top) Cortical seed-based FC for the Dorsal Area 24dd Glasser ROI (black region, top image) estimated from deconvolved neural signals using both our two-stage approach and rsHRF for test-retest scans from a randomly selected HCP subject (103818). The bottom row compares results of the two approaches on a sample vertex level BOLD signal in the ROI.}}
    \label{fig:seed_based_FC_surface_plot}
\end{figure}
\par
\revised{\textbf{Functional Connectivity}:} \revised{Figure~\ref{fig:seed_fc_dbICC} shows a boxplot of the seed-based left-hemisphere FC dbICCs for all Glasser atlas parcels for both our method and rsHRF, estimated using all $45$ HCP test-retest subjects. Our method exhibits significantly higher FC reproducibility than rsHRF (parcel-paired $t$-test p-value $<10^{-9}$), albeit the improvement is not as dramatic as that observed for HRF reproducibility in Table~\eqref{tab:hrf_reproducibility}. For qualitative evaluation, Figure~\ref{fig:seed_based_FC_surface_plot} shows the seed-based FC for Glasser area 24dd in test-retest subject 103818. The spatial FC patterns are broadly consistent across scans, similar between methods, and are well aligned with the expected functional organization reported in the literature (see \cite{baker2018connectomic}. Figure 5), although rsHRF estimates correlations of greater magnitude. The bottom panels in Figure~\ref{fig:seed_based_FC_surface_plot} show the deconvolved signals, estimated HRF and reconstructed BOLD signals for both methods at the same representative vertex within the seed region over a $100$-s time window. The estimated neural signals differ substantially between methods, which is expected, as the different approaches to deconvolution are making different prior assumptions on the nature of the underlying neural signals, namely sparsity (Ours) versus an optimal linear filter (rsHRF). While the improved reproducibility of the proposed approach is promising, the absence of ground truth of course precludes determining whether this improvement corresponds to more accurate recovery of the underlying neural activity or hemodynamic coupling.}
\subsubsection{Population Analysis}

\begin{figure}[!ht]
  \centering
    \includegraphics[width=\textwidth]{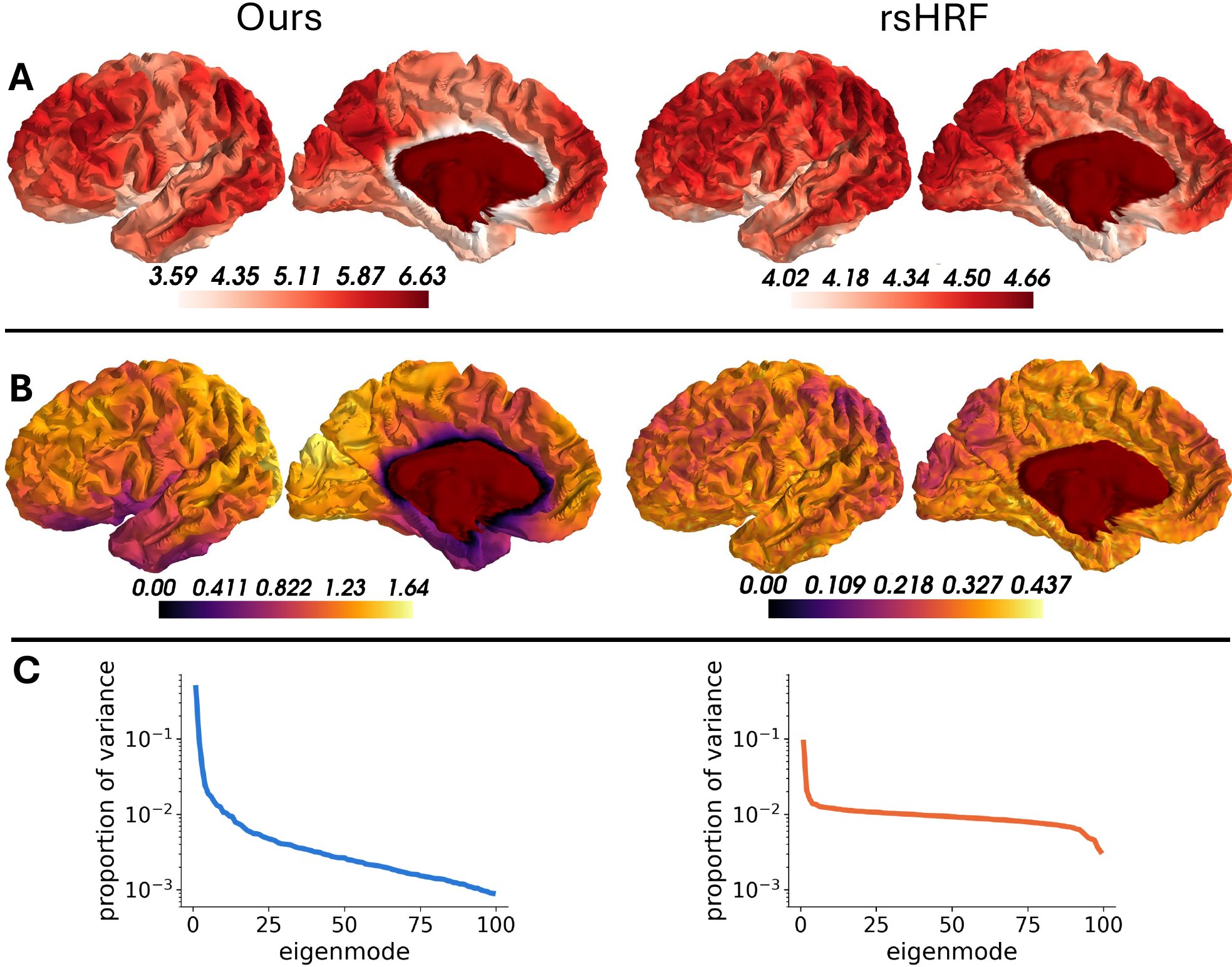}
    \caption{\revised{A) Mean Time-to-peak (in seconds) fields,  B) pointwise standard deviation of time-to-peak fields $l(x)$, C) eigenvalue spectral decay, estimated from $100$ unrelated HCP-YA subjects using both our method and rsHRF.}}
    \label{fig:mean_var_spectraTTP}
\end{figure}

\begin{figure}[!ht]
  \centering
    \includegraphics[width=\textwidth]{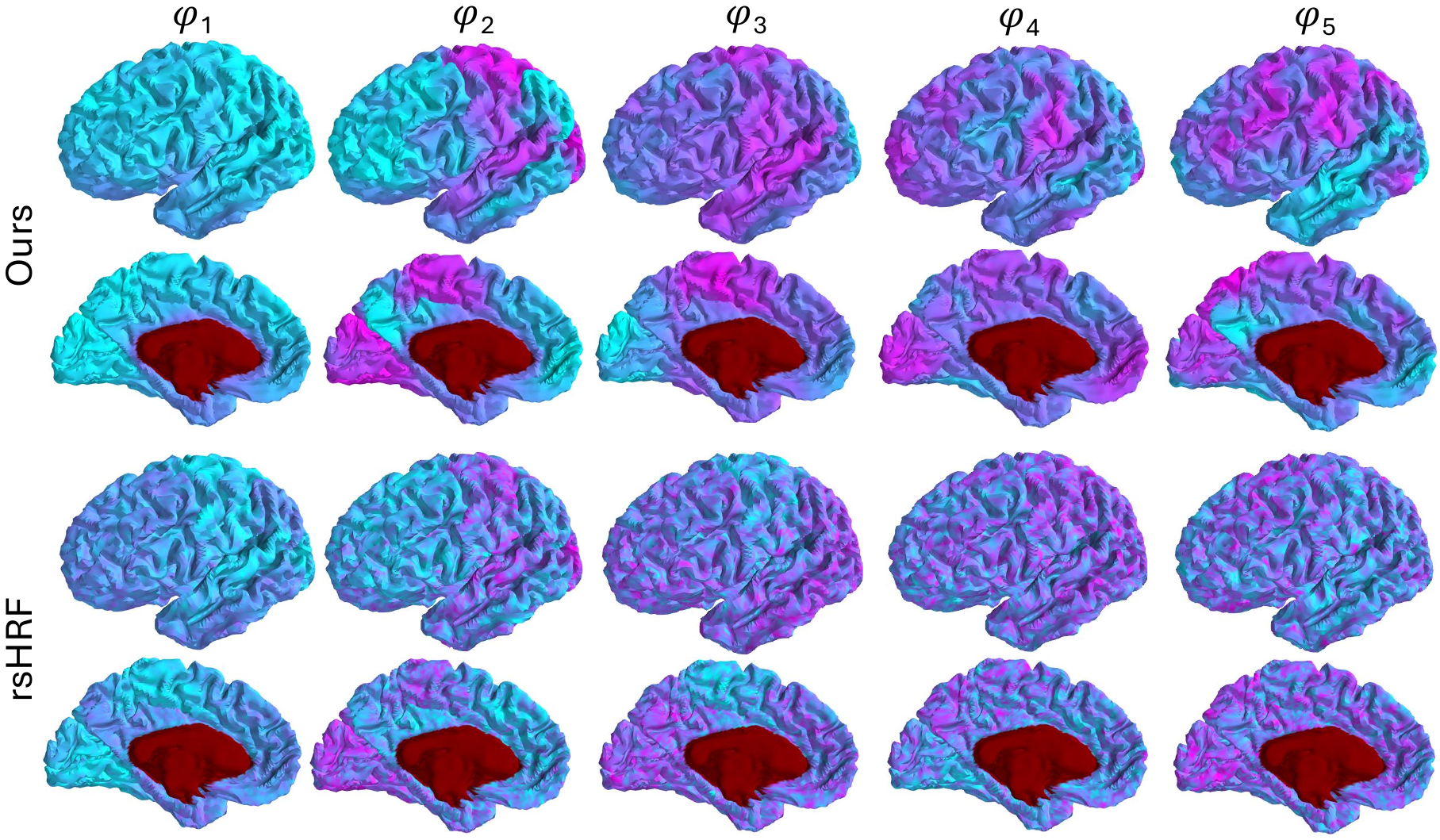}
    \caption{\revised{The first $5$ eigenfunctions $\varphi_k$, describing the main modes of inter-subject variability, estimated from $100$ unrelated HCP-YA subjects using both our method and rsHRF.}}
    \label{fig:eigfuncsTTP}
\end{figure}

\revised{Panel A of Figure \ref{fig:mean_var_spectraTTP} shows the mean time-to-peak (in seconds) of the HRF across the cortical surface computed from the hemodynamic coupling estimated using both our method and rsHRF. We note both methods estimate less mean latency in several of the sensory and motor areas (pre/postcentral gyrus), along with some regions in the inferior temporal lobe. The inferior temporal lobe is often affected by susceptibility-induced $B_0$ inhomogeneities that can cause signal dropout \citep{WEISKOPF2006493}, and therefore the reduced latency estimates may partly reflect data quality issues rather than physiology. Areas with more latency include regions in the frontal and inferior parietal lobes. 
Excluding the boundary region near the medial wall, where our estimates exhibit some edge effects stemming from the spatial regularization, the proposed method yields mean time-to-peak estimates ranging from approximately $4$ to $6.6$ seconds, which is consistent with ranges reported in the literature \citep{handwerker2004variation}. In contrast, the mean estimates obtained with rsHRF span a substantially narrower range of approximately $4$ to $4.6$ seconds.} 
\revised{Panel B of Figure~\ref{fig:mean_var_spectraTTP}  shows the point-wise standard deviation of the time-to-peak fields estimated using the low-rank approximation \eqref{eqn:low_rank_cov} across the HCP unrelated sample. We see that the time-to-peak estimates from our method exhibit large variances in areas of the frontal lobe and inferior parietal lobe, indicating substantial inter-subject variability. This is consistent with the subject-level results in Figure~\ref{fig:reproducibility}, where the frontal-lobe hemodynamics of the topmost subject differ substantially from those of the other two subjects, highlighting the well-documented inter-subject hemodynamic variability \citep{lindquist2009modeling}. When comparing the point-wise standard deviations obtained using the two hemodynamic estimation methods, we observe two notable differences. First, in contrast to the mean, the shape of spatial variability differs substantially between the methods. Second, the time-to-peak estimates from our method exhibit a larger overall magnitude of pointwise variability than those from rsHRF, suggesting that our method estimates greater population-level heterogeneity in hemodynamic latency.}
\par 
\revised{Figure~\ref{fig:eigfuncsTTP} shows the first five eigenfunctions computed by applying FPCA to the time-to-peak estimates from our method (top two rows) and rsHRF (bottom two rows) using the HCP unrelated sample. For both methods, the eigenfunctions become progressively ``rougher'' and exhibit higher-frequency spatial variation with increasing mode, as expected. However, the eigenfunctions computed from the proposed method exhibit much more spatial regularity than those from rsHRF, with smooth, local structures observed even in the higher modes. Panel C of Figure~\ref{fig:mean_var_spectraTTP} shows the proportion of variance explained by each eigenmode. We see that the leading eigenfunctions from our method capture a much higher proportion of the total sample variance than those from rsHRF. Taken together, these results indicates that, using the latency estimates from our our method, a larger fraction of the estimated between-subject variability is concentrated in a smaller number of spatial modes which exhibit substantially greater spatial regularity and smoothness.}

\section{Conclusion and Discussion}\label{sec:discussion_conclusion}
In this work, we present a new methodology for estimating a spatially varying, subject-specific forward model of hemodynamic coupling from resting-state fMRI. To disentangle estimation of the latent neural signal from that of the forward hemodynamic model, we adopt a marginal likelihood approach in which the latent signal is marginalized out. The resulting objective function is approximated using a deep conditional density estimator. Spatial regularization is enforced via a latent Gaussian process prior, and estimation uncertainty is quantified using a double-bootstrap procedure. Given the estimated forward model, latent neural signals can then be recovered using any standard non-blind inference method. In both synthetic and in-vivo datasets, our approach results in more accurate \revised{and reproducible} hemodynamic estimates and improves downstream \revised{functional connectivity reproducibility} versus standard competing methods. 
\par 
A thus far understated feature of our method is computational efficiency. By exploiting the sparsity of the prior precision over a finite element basis, estimation can be performed on ultra high resolution meshes using off-the-shelf algorithms to solve the sparse linear system \eqref{eqn:NewtonLinSystem}. For instance, point estimation on the $32,492$ vertex surface mesh took on average around 30 seconds for model \eqref{eqn:double_gamma_hrf} and about 150 seconds for model \eqref{eqn:canonical_gamma_basis}, on a personal CPU machine with 36GB of RAM and 12 cores. Note that while these runtimes already make this method practical for practitioners to run on standard modern laptops, it is highly likely they could be (significantly) improved by migrating the code to a GPU, which would speed both the backpropagation used to compute the derivatives in \eqref{eqn:logPostGrad} and \eqref{eqn:logPostHess} 
as well as the preconditioned conjugate gradient solves by leveraging existing GPU-optimized implementations.
\par 
This work has a few limitations. \revised{First,} while empirical results on both synthetic and in-vivo data indicate good calibration of the double bootstrap algorithm for uncertainty quantification, a limitation of this work is the computational burden of this procedure, as it requires $BR$ point estimates. In practice, we use $B=100$ and \revised{$R=100$}. However, each of the outer bootstraps is independent and therefore can be parallelized, provided sufficient computational resources. 
\par 
\revised{In our experiments, we evaluate undirected functional connectivity of the deconvolved neural signals under the estimated hemodynamic forward model using a seed-based reproducibility analysis, demonstrating improvements over the standard approach. However, in certain applications, directed (effective) connectivity may be of more interest. Effective connectivity is often quantified using multivariate Granger-causality measures. However, such approaches scale poorly with the dimension of the time series, with computational complexity on the order $V^2$ \citep{shojaie2022granger}, making application to the surface-based mesh deconvoled signals of dimension $V=32,492$ impractical. Integrating the proposed framework with effective connectivity estimation is therefore non-trivial, and would likely require some form of dimension reduction.}
\par 
\revised{The learned marginal likelihood in Section~\ref{ssec:learnedLik} requires specifying a simulator for the neural signals, hemodynamic response, and measurement noise. Due to the complexity of the biophysics, no BOLD simulator can be expected to exactly reproduce the true data-generating process, and therefore some degree of model misspecification is likely unavoidable. The synthetic experiments in Section~\ref{ssec:synth_results} demonstrate robustness of the proposed method to some forms of misspecification, but many other forms may arise in practice, including both unmodeled biophysical processes and deviations from the assumed white-noise measurement process. It is well known that such simulator misspecification can introduce varying degrees of bias in parameter estimates \citep{brynjarsdottir2014learning}, and so a principled manner of evaluating and accommodating this within our learned marginal likelihood framework is of interest.}
\par 
Our method can be extended \revised{and applied} in several ways. \revised{The posterior mean summary statistic~\eqref{eqn:posterior_mean_summary}was chosen for its guaranteed dimension reduction and simple $\ell_2$-based training objective. However, because it is not generally sufficient, it may discard information relevant for inference in some settings. An important direction for future work is to investigate alternative summary statistics and joint representation-likelihood learning approaches \citep{dirmeier2025}, and to evaluate their impact on the overall inversion.}
In our experiments, we use a classical deconvolution \revised{method, outlined in Section~\ref{ssec:signal_deconv},} to estimate the neural signals under the corrected hemodynamic forward model independently at each vertex. In reality, there exists a complex dependence structure in the neural signals, mediated by both local and non-local white matter fiber tracts. Designing a high-resolution signal inverter that is able to effectively leverage this correlation is of significant interest. 
\par 
\revised{On the application side, an important question is which hemodynamic coupling parameters are practically identifiable under a given TR, signal-to-noise ratio, and run length $M$. Under an HCP-like protocol, the experiments in Section~\ref{ssec:synth_results} provide substantially stronger empirical validation for the one-parameter HRF model~\eqref{eqn:double_gamma_hrf}, as $\theta_2$ in the two-parameter model~\eqref{eqn:canonical_gamma_basis} exhibits poor practical identifiability, reflected in the wide confidence intervals produced by the resampling-based UQ procedure in Section~\ref{ssec:uq}. More generally, the proposed UQ procedure provides a practical tool for assessing parameter identifiability across acquisition settings and may help inform both model specification and acquisition protocol design.
Finally, there is significant interest in the potential of hemodynamic coupling as a biomarker, including in neuropsychiatric disorders \citep{Wenjing2021} and aging \citep{baron2025revealing}. Applying the proposed high-resolution hemodynamic coupling estimates to population-level statistical analyses is an important direction for future work.}


\bibliographystyle{chicago}
\bibliography{refs}

\newpage 
\begin{center}
    \Large \textbf{Supplemental Material to ``Learned Hemodynamic Coupling Inference in Resting-State Functional MRI''}
\end{center}

\setcounter{equation}{0}
\setcounter{figure}{0}
\setcounter{table}{0}
\setcounter{section}{0}
\setcounter{page}{1}
\makeatletter
\renewcommand{\theequation}{S\arabic{equation}}
\renewcommand{\thefigure}{S\arabic{figure}}
\renewcommand{\thetable}{S\arabic{table}}
\renewcommand{\thesection}{S\arabic{section}}
\renewcommand{\bibnumfmt}[1]{[S#1]}
\renewcommand{\citenumfont}[1]{S#1}
\renewcommand{\theequation}{S.\arabic{equation}}
\renewcommand{\thesection}{S\arabic{section}}
\renewcommand{\thesubsection}{S\arabic{section}.\arabic{subsection}}
\renewcommand{\thetable}{S\arabic{table}}
\renewcommand{\thefigure}{S\arabic{figure}}

\section{Derivation of Discretized Hemodynamic Coupling Field Prior}\label{supp:derivations}
We form a finite element approximation of the $j$'th coordinate field
\begin{equation}\label{eqn:hemo_field_finite_elements_basis}
    \tilde{\theta}_j(x)=\sum_{v=1}^{V}w_{jv}\phi_v(x),
\end{equation}
where $w_j=(w_{j1},\ldots,w_{jV})^\intercal$ is a (random) basis coefficient vector. The weak formulation of \eqref{eqn:MaternSPDE} over the finite-dimensional function space $\text{span}\{\{\phi_{v}\}_{v=1}^V\}$ with $\beta_j=2$ implies 
\begin{equation}\label{eqn:WeakMaternSPDEBasis}
    \langle -(\Delta_{\Omega} - \kappa_j^2I)\tilde{\theta}_j, \phi_v\rangle = -\langle \Delta_{\Omega}\tilde{\theta}_j,\phi_v \rangle  + \kappa_j^2\langle \tilde{\theta}_j, \phi_v\rangle = \frac{1}{\tau_j}\langle u, \phi_v\rangle, \quad v=1,...,V,
\end{equation}
where $\langle f,g\rangle := \int_{\Omega}fg$, i.e. $L^2(\Omega)$ inner product. Apply integration by parts to the inner product involving $\Delta_{\Omega}$, we can write \eqref{eqn:WeakMaternSPDEBasis} as  
\begin{equation}\label{eqn:WeakMaternSPDEBasisIBP}
\left\langle \nabla_{\Omega}\phi_v,\nabla_{\Omega}\tilde{\theta}_j\right\rangle
+\kappa_j^2\left\langle \phi_v,\tilde{\theta}_j\right\rangle
=
\frac{1}{\tau_j}\left\langle \phi_v,u\right\rangle, \quad v=1,...,V.
\end{equation}
Now, plugging in the basis expansion \eqref{eqn:hemo_field_finite_elements_basis} for $\tilde{\theta}_j$ into \eqref{eqn:WeakMaternSPDEBasisIBP} and recalling the definition of inner product matrices of $C$ and $G$ from \eqref{eqn:mass_stiffness_matrices}, we get the vectorized system of (random) linear equations  
$$
\left[G + \kappa_j^2 C\right]w_j
=
\frac{1}{\tau_j}\xi,
$$
where $\xi\in\mathbb{R}^V$ is a random vector with elements $\xi_v=\langle \phi_v,u\rangle$. Since $u$ is Gaussian white noise, by properties of Gaussian measures the $\xi_v$'s are normal random variables with mean 
$$
\mathbb{E}\left[\langle \phi_v,u\rangle\right] = 0, \quad \text{Cov}\left(\xi_v,\xi_{v'}\right) = \mathbb{E}\left[\langle \phi_v,u\rangle\langle \phi_{v'},u\rangle\right] = \langle \phi_v, \phi_{v'}\rangle := C_{v,v'},
$$
hence
$$
\xi \sim\mathcal{N}(0,C).
$$
It follows that
$$
w_j
=
\frac{1}{\tau_j}\left[\kappa_j^2 C + G\right]^{-1}\xi,
$$
and by standard properties of Gaussians we have 
$$
w_j
\sim
\mathcal{N}\left(
0,
\frac{1}{\tau_j^2}
\left[\kappa_j^2 C + G\right]^{-1} C \left[\kappa_j^2 C + G\right]^{-1}\right).
$$
Hence, the corresponding precision matrix is
$$
Q_j
=
\tau_j^2
\left(\kappa_j^2C+G\right)^\intercal
C^{-1}
\left(\kappa_j^2C+G\right).
$$
The desired result is obtained using the nodal property of the finite element basis, that is, at any mesh point $x_v$, the representation \eqref{eqn:hemo_field_finite_elements_basis} results in 
$$
\tilde{\theta}_j(x_v) = w_{jv},
$$
hence  
$
\tilde{\boldsymbol{\theta}}_j = w_j \sim \mathcal{N}\left(
0,Q_j^{-1}\right)$.

\end{document}